\documentclass[fleqn,usenatbib]{mnras}

\usepackage{newtxtext,newtxmath}

\usepackage[T1]{fontenc}

\DeclareRobustCommand{\VAN}[3]{#2}
\let\VANthebibliography\thebibliography
\def\thebibliography{\DeclareRobustCommand{\VAN}[3]{##3}\VANthebibliography}

\usepackage{graphicx}	% Including figure files
\usepackage{amsmath}	% Advanced maths commands
\usepackage{bm}
\usepackage{booktabs}
\usepackage{threeparttable}
\usepackage{multicol}
\usepackage{xcolor}
\usepackage[normalem]{ulem}
\usepackage[export]{adjustbox}
\usepackage{pdflscape}
\usepackage{soul}

\newcommand{\email}[1]{\mbox{\href{mailto:#1}{#1}}}

\makeatletter
\newlength{\abovecaptionskip}%
\makeatother

\defcitealias{mandel22}{M20}
\defcitealias{ward22}{W22}

\DeclareMathOperator*{\argmin}{arg\,min}
\DeclareMathOperator{\exponential}{Exponential}
\DeclareMathOperator{\halfcauchy}{Half-Cauchy}

\allowdisplaybreaks

\title[Dust Laws from the RAISIN Survey]{Using Rest-Frame Optical and NIR Data from the RAISIN Survey to Explore the Redshift Evolution of Dust Laws in SN Ia Host Galaxies}

\author[S.\ Thorp et al.]{Stephen Thorp$^{1,2}$\thanks{E-mail: \email{stephen.thorp@fysik.su.se}}, Kaisey S.\ Mandel$^{2,3}$, David O.\ Jones$^{4}$, Robert P.\ Kirshner$^{5,6}$, and Peter M.\ Challis$^{7}$\\
$^1$The Oskar Klein Centre, Department of Physics, Stockholm University, AlbaNova University Centre, SE 106 91 Stockholm, Sweden\\
$^2$Institute of Astronomy and Kavli Institute for Cosmology, Madingley Road, Cambridge, CB3 0HA, UK\\
$^3$Statistical Laboratory, DPMMS, University of Cambridge, Wilberforce Road, Cambridge, CB3 0WB, UK\\
$^4$Institute for Astronomy, University of Hawai'i, 640 N.\ A'ohoku Pl., Hilo, HI 96720, USA\\
$^5$TMT International Observatory, 100 West Walnut Street, Pasadena, CA 91124, USA\\
$^6$California Institute of Technology, 1200 East California Boulevard, Pasadena, CA 91125, USA\\
$^7$Harvard--Smithsonian Center for Astrophysics, 60 Garden Street, Cambridge, MA 02138, USA 
}

\date{Accepted XXX. Received YYY; in original form ZZZ}

\pubyear{2024}

\begin{document}
\label{firstpage}
\pagerange{\pageref{firstpage}--\pageref{lastpage}}
\maketitle

% Abstract of the paper
\begin{abstract}
%It should be a single paragraph not more than 250 words (200 words for Letters). No references should appear in the abstract.
We use rest-frame optical and near-infrared (NIR) observations of 42 Type Ia supernovae (SNe Ia) from the Carnegie Supernova Project at low-$z$ and 37 from the RAISIN Survey at high-$z$ to investigate correlations between SN Ia host galaxy dust, host mass, and redshift. This is the first time the SN Ia host galaxy dust extinction law at high-$z$ has been estimated using combined optical and rest-frame NIR data ($YJ$-band). We use the \textsc{BayeSN} hierarchical model to leverage the data's wide rest-frame wavelength range (extending to $\sim1.0$--1.2~$\upmu$m for the RAISIN sample at $0.2\lesssim z\lesssim0.6$). By contrasting the RAISIN and CSP data, we constrain the population distributions of the host dust $R_V$ parameter for both redshift ranges. We place a limit on the difference in population mean $R_V$ between RAISIN and CSP of $-1.16<\Delta\mu(R_V)<1.38$ with 95\% posterior probability. For RAISIN we estimate $\mu(R_V)=2.58\pm0.57$, and constrain the population standard deviation to $\sigma(R_V)<0.90~[2.42]$ at the 68~[95]\% level. Given that we are only able to constrain the size of the low- to high-$z$ shift in $\mu(R_V)$ to $\lesssim1.4$ -- which could still propagate to a substantial bias in the equation of state parameter $w$ -- these and other recent results motivate continued effort to obtain rest-frame NIR data at low and high redshifts (e.g.\ using the \textit{Roman Space Telescope}). 
\end{abstract}
%243

% Select between one and six entries from the list of approved keywords.
% Don't make up new ones.

%%%%%%%%%%%%%%%%%%%%%%%%%%%%%%%%%%%%%%%%%%%%%%%%%%

%%%%%%%%%%%%%%%%% BODY OF PAPER %%%%%%%%%%%%%%%%%%

\section{Introduction}
With the advent of the Vera C. Rubin Observatory's Legacy Survey of Space and Time \citep[LSST;][]{ivezic19}, and the High-Latitude Time Domain Survey (HLTDS) on the \textit{Nancy Grace Roman Space Telescope} \citep{spergel15, hounsell18, rose21}, cosmology using Type Ia Supernovae (SNe Ia) is set to enter an era of unprecedented precision. However, the success of these future experiments hinges on our ability to account for systematic uncertainties that are not yet fully understood. A particularly challenging and contentious issue is determining the distribution of dust laws in SN Ia host galaxies \citep[see e.g.][]{brout21, thorp21}. Galaxy evolution studies tell us that dust properties correlate strongly with stellar mass and star formation \citep[see e.g.][]{salim18, nagaraj22}, but the parent stellar populations of SNe Ia also correlate with galaxy properties \citep[e.g.][]{childress14}. Either of these effects could cause the accuracy of SN Ia distance estimates to ``drift'' over cosmic time -- a deeply troubling systematic for cosmology. The very latest constraints on the dark energy equation-of-state parameter ($w$; see \citealp{des24}) from SNe Ia have highlighted the importance of understanding this issue \citep{vincenzi24}. In this paper, we analyse new rest frame near-infrared (NIR) observations of SNe Ia at high redshift from the RAISIN Survey \citep[SNIA in the IR;][]{jones22}, using these data in combination with the \textsc{BayeSN} hierarchical model \citep{mandel22, grayling24} to explore how the dust laws in SN Ia host galaxies depends on redshift and mass.

The nature of line-of-sight dust extinction in SN Ia host galaxies has been a topic of significant investigation throughout the history of supernova cosmology, with the correction of this effect being a critical part of SN Ia standardisation. A particular sticking point has been the estimation of the dust law $R_V$ parameter in SN Ia hosts -- an issue debated in the literature \citep[e.g.][and references therein]{branch92, riess96, tripp98, tripp99} since before the first discovery of accelerating expansion \citep{riess98, perlmutter99}. As acknowledged in many of these early works \citep[particularly][]{riess96}, the estimation of $R_V$ is rendered challenging by the confounding between this quantity and any intrinsic colour--luminosity correlation exhibited by SNe Ia \citep[see][for extensive discussion of this problem]{mandel17}. Considerable progress has been made over the past two and a half decades, with the construction of large optical+NIR SN Ia samples \citep[e.g.][]{woodvasey08, contreras10, stritzinger11, friedman15, krisciunas17}, and the development of robust statistical methods for analysing these \citep[e.g.][]{mandel09, mandel11, burns11, burns14, mandel22}.

Nevertheless, challenges and uncertainty persist. As well as continued debate over whether SN Ia hosts are consistent with Milky Way-like dust ($R_V\approx3$ with small variation around this; see e.g.\ \citealp{schlafly16}), there is ongoing discussion regarding the level of $R_V$ variation amongst SN Ia host galaxies, and the level of correlation between this and galaxy mass (see e.g.\ \citealp{brout21, thorp21, johansson21, gonzalezgaitan21, wiseman22, wiseman23, duarte22, meldorf22, kelsey22, popovic23, karchev23_sim, karchev24, grayling24, wojtak24}). We review the past three years of $R_V$ estimates in Appendix \ref{app:rv}. There is a well known correlation between post-standardization SN Ia brightness (or Hubble residuals) and host mass -- often referred to as a ``mass step'' \citep[see e.g.][and many others]{kelly10, sullivan10}. A difference in host galaxy dust law $R_V$ between low- and high-mass host galaxies has been proposed by \citet{brout21} as an explanation for this effect\footnote{It is worth noting that several much earlier studies \citep{sullivan10, lampeitl10} had reported a significant difference in $R_V$ or colour--luminosity slope, $\beta$, between passive and star forming SN host galaxies. However, \citet{lampeitl10} did not make a direct link between this and the mass step they reported, whilst \citet{sullivan10} strongly favoured an interpretation of the mass step based on intrinsic SN Ia properties (particularly progenitor metallicity, \`a la \citealp{timmes03}), as they found similar apparent colours in low- and high-mass hosts.}. Studies of dust attenuation laws in Dark Energy Survey (DES) SN Ia host galaxies have partially supported this picture \citep{meldorf22, duarte22}. Hubble residual steps as a function of host mass and host $U-R$ colour (the latter being a tracer of stellar population age) have been studied by \citet{wiseman22} for the DES 5~yr SN Ia sample. who found it difficult to reconcile this with a solely dust-based explanation \citep[see also][]{kelsey22, wiseman23}. They suggest that a residual effect relating to progenitor metallicity \citep[see e.g.][]{hoflich98, timmes03} may also be at play\footnote{The possible cosmological impact of such a metallicity effect has been considered since the original discovery of accelerating expansion \citep{hoflich98,riess98}.}. The latest DES analyses \citep{vincenzi24, des24} have supported the findings of \citet{wiseman22}, and concluded that a completely dust-based mass step model \citep[\`a la][]{brout21} cannot fully explain their data.

Tentative detections of a mass step at NIR wavelengths (where sensitivity to dust should be much lower) would also seem to contradict a scenario where dust is the only driver of SN--host correlations \citep[see][]{uddin20, ponder21, johansson21, jones22, uddin23, peterson24}. Previous hierarchical Bayesian analyses of SN Ia light curves using \textsc{BayeSN} \citep{thorp21, thorp22} did not find a significant difference in line-of-sight $R_V$ as a function of host galaxy stellar mass bins, and found that a non-zero residual brightness step is still required to explain the data, even when allowing for different $R_V$ distributions in low- and high-mass hosts. Very recently \citet{grayling24} applied \textsc{BayeSN} to a large set of optical data from Foundation \citep{foley18, jones19}, Pan-STARRS \citep{scolnic18}, and DES \citep{brout19_des}. They explored a variety of models, including one where SNe in low- and high-mass hosts are allowed to differ at the level of the intrinsic SED. They concluded that both intrinsic and extrinsic effects are likely at the root of the mass step. A study of the host-mass dependence of the SALT3 spectroscopic model \citep{jones23} found that $\approx35$ per cent of the mass step is due to spectral variations that are independent of luminosity (i.e.\ not easily explainable by dust). A hierarchical Bayesian analysis by \citet{wojtak23} applied a two-population model without an explicit connection to host galaxy mass. Their results showed a preference for two populations that differ in intrinsic colour, light curve stretch, and dust properties. Studies of early time SN Ia behaviour \citep[e.g.][]{ye24} have also hinted at intrinsic diversity in the SN Ia population.

Correlations of SNe Ia Hubble residuals and light curve shapes with host galaxy mass or star formation rate have frequently been linked \citep[see e.g.][]{sullivan06,sullivan10, childress13, childress14, kim18, rigault20, nicolas21, wiseman22, briday22, chung23} to models where SNe arise from a ``prompt'' population of young progenitors, and a ``delayed'' population of older progenitors (as proposed by \citealp{mannucci05, scannapieco05, mannucci06}). Under such a model, the evolution of galaxy properties across cosmic time could cause an evolution of the progenitor distribution, and thus an evolving ``mass step'' \citep[see e.g.][]{sullivan10, rigault13, childress14}. This could be a significant source of systematic uncertainty for current and future dark energy studies.

A dependence of host galaxy dust properties on host mass or star formation rate could equally give rise to problematic cosmological systematics. It would create a ``$\beta$-evolution'' effect \citep{kessler09, sullivan10, conley11}, where the appropriate SN Ia colour--luminosity corrections, or correct treatment of dust, evolve with redshift ($R_V$-evolution is discussed specifically by \citealp{nordin08}). Although Pan-STARRS, Pantheon, and the DES 3~yr analyses had not seen strong evidence for such an effect \citep{jones18, scolnic18, brout19}, the very recent analysis of the 5~yr DES dataset claims some evidence for redshift evolution of $\beta$ \citep{vincenzi24, des24}. Since these DES 5~yr results have shown a non-significant preference for ``quintessence'' ($-1<w<0$; \citealp{caldwell98}) over a cosmological constant ($w=-1$), it is crucial to understand any systematic uncertainties in our analyses.

As discussed above, there has been extensive recent investigation of whether dust law $R_V$ values in SN host galaxies correlate with galaxy stellar mass \citep[e.g.][]{brout21, thorp21, johansson21, meldorf22, thorp22, vincenzi24, grayling24}. However, less investigated is the possibility of correlations between $R_V$ and redshift, either in addition to, or instead of, galaxy stellar mass. \citet{grayling24} recently presented one of the first analyses of this kind, fitting (to Foundation, Pan-STARRS, and DES data) a model where the mean of the $R_V$ population distribution evolves linearly with redshift. They estimated that the gradient of this relation is $\eta_R=-0.38\pm0.70$ (where negative values would imply a mean $R_V$ that decreases with redshift). It is not yet clear if claimed results from SN Ia host galaxies are well aligned with studies of galaxies more generally \citep[see discussion in][]{wiseman22, meldorf22, duarte22}, since galaxy photometry constrains the attenuation law rather than extinction law \citep[for a review see][]{salim20}. Nevertheless, galaxy evolution studies suggest there are significant correlations between dust attenuation, galaxy stellar mass, star formation rate, and redshift \citep[e.g.][]{garn10, zahid13, salim18, salim20, nagaraj22, alsing24}. Independently of any correlations with host galaxies, a redshift-dependence of the dust law $R_V$ affecting SNe could also arise from intergalactic dust, although such an effect is likely to be small \citep[see e.g.][]{mortsell03, menard10_qsos, menard10_sne, johansson12}.

Recently, the RAISIN Survey \citep{jones22} obtained rest-frame NIR observations of 37 higher-redshift SNe Ia ($0.2 \lesssim z \lesssim 0.6$) using the \textit{Hubble Space Telescope} (\textit{HST}). Previously, a large sample of SNe Ia in this redshift range had only been probed out to the rest frame $I$-band \citep{nobili05, nobili09, freedman09}, although redder wavelengths were observed for a small number of supernovae by \citet{stanishev18}. Observations of individual lensed SNe Ia have also probed the rest frame NIR at high-$z$ \citep[e.g.][]{quimby13, dhawan20, goobar23, pierel24}. Given the advantages of the NIR for constraining the properties of host galaxy dust \citep[see e.g.][]{krisciunas07, mandel11, mandel22, thorp22}, the RAISIN data thus present us with an excellent and unprecedented opportunity to estimate the host galaxy $R_V$ values of SNe Ia at higher-redshift, to test if these differ strongly from those estimated at lower-$z$. Therefore, in this paper we apply our \textsc{BayeSN} hierarchical model \citep{mandel22} to the RAISIN data to investigate precisely this question. We follow the analysis scheme demonstrated by \citet{thorp22} to constrain the distribution of dust law $R_V$ values in the higher-$z$ RAISIN sample, and the lower-$z$ Carnegie Supernova Project companion sample analysed by \citet{jones22}. As well as estimating the $R_V$ distributions in these two samples, we also consider the effect of host galaxy stellar mass. This work is intended to serve as a complement to the recent \textsc{BayeSN} analysis by \citet{grayling24} of a large optical dataset, which also investigated correlations between $R_V$, redshift and host galaxy mass.

In Section \ref{sec:data}, we describe the SN Ia samples that we are analysing. In Section \ref{sec:model}, we recap the key details of the \textsc{BayeSN} model, and the mode of analysis that we will follow. We discuss our results in Section \ref{sec:results}, and finally provide our conclusions in Section \ref{sec:conclusions}.

\section{Data}
\label{sec:data}
We use the same low- and high-redshift SN Ia samples as in the RAISIN cosmology analysis \citep{jones22}. At low redshift ($z<0.1$), we use the optical and NIR photometry of 42 SNe Ia from the Carnegie Supernova Project \citep[CSP;][]{krisciunas17}. These are chosen to all have redshifts of $z>0.01$ to avoid excessive uncertainty due to peculiar velocity. The CSP supernovae were selected for follow-up from galaxy-targeted surveys (i.e.\ surveys monitoring a fixed list of galaxies). At high redshift ($0.22\leq z\leq0.61$), we use data for 37 SNe Ia from the RAISIN Survey, as presented by \citet{jones22}. All of these have NIR (\textit{F125W} and \textit{F160W}) photometry from the \textit{Hubble Space Telescope}. For 19 of the SNe from RAISIN1, these data are augmented by optical ($griz$) data from the Pan-STARRS Medium Deep Survey \citep[MDS;][]{chambers16,scolnic18,villar20}. The 18 remaining SNe from RAISIN2 have optical ($griz$) data from the Dark Energy Survey \citep{des16,abbott19,brout19}. The SNe in both components of the sample are selected to have low--moderate host reddening ($E(B-V)<0.3$, consistent with the cuts used in typical cosmological analyses), with no excessively fast or slow decliners (\textsc{SNooPy} colour-stretch parameter in the range $0.75<s_{BV}<1.18$). The SNe in the final sample have all been identified as spectroscopically normal SNe Ia.

We use the same redshifts as in \citet{jones22}, whose peculiar velocity corrections are based on the 2M++ catalogue \citep{lavaux11}. We compute external distance estimates, $\hat{\mu}_{\text{ext},s}$, from these, using the distance--redshift relation one would find under a flat $\Lambda$CDM cosmology (with an assumed $H_0=73.24$~km\,s$^{-1}$\,Mpc$^{-1}$, $\Omega_M=0.28$, and $\Omega_\Lambda=0.72$; \citealp{riess16}). Our uncertainties on our external distance estimates are computed following equation 27 of \citet{mandel22}, based on contributions from the spectroscopic redshift uncertainty and an assumed peculiar velocity uncertainty of 150~km\,s$^{-1}$\,Mpc$^{-1}$ \citep{carrick15}. Milky Way extinction is corrected using the \citet{schlafly11} catalogue. Host galaxy mass estimates are determined in a self consistent way for all SNe in the sample using LePhare \citep{arnouts11}, as described in \citet[appendix D]{jones22}. The CSP sample is weighted towards high masses, with $35/42$ ($24/42$) SNe having estimated host galaxy stellar masses above $10^{10}~\mathrm{M}_\odot$ ($10^{10.44}~\mathrm{M}_\odot$)\footnote{As discussed in \citet{jones22}, the alternative step location of $10^{10.44}~\mathrm{M}_\odot$ is chosen as a cross-check, based on \citet{ponder21}.}. For the higher-$z$ RAISIN sample the reverse is true, with $11/37$ ($6/37$) SNe having estimated host masses greater than $10^{10}~\mathrm{M}_\odot$ ($10^{10.44}~\mathrm{M}_\odot$).

\begin{figure}
    \centering
    \includegraphics[width=\linewidth]{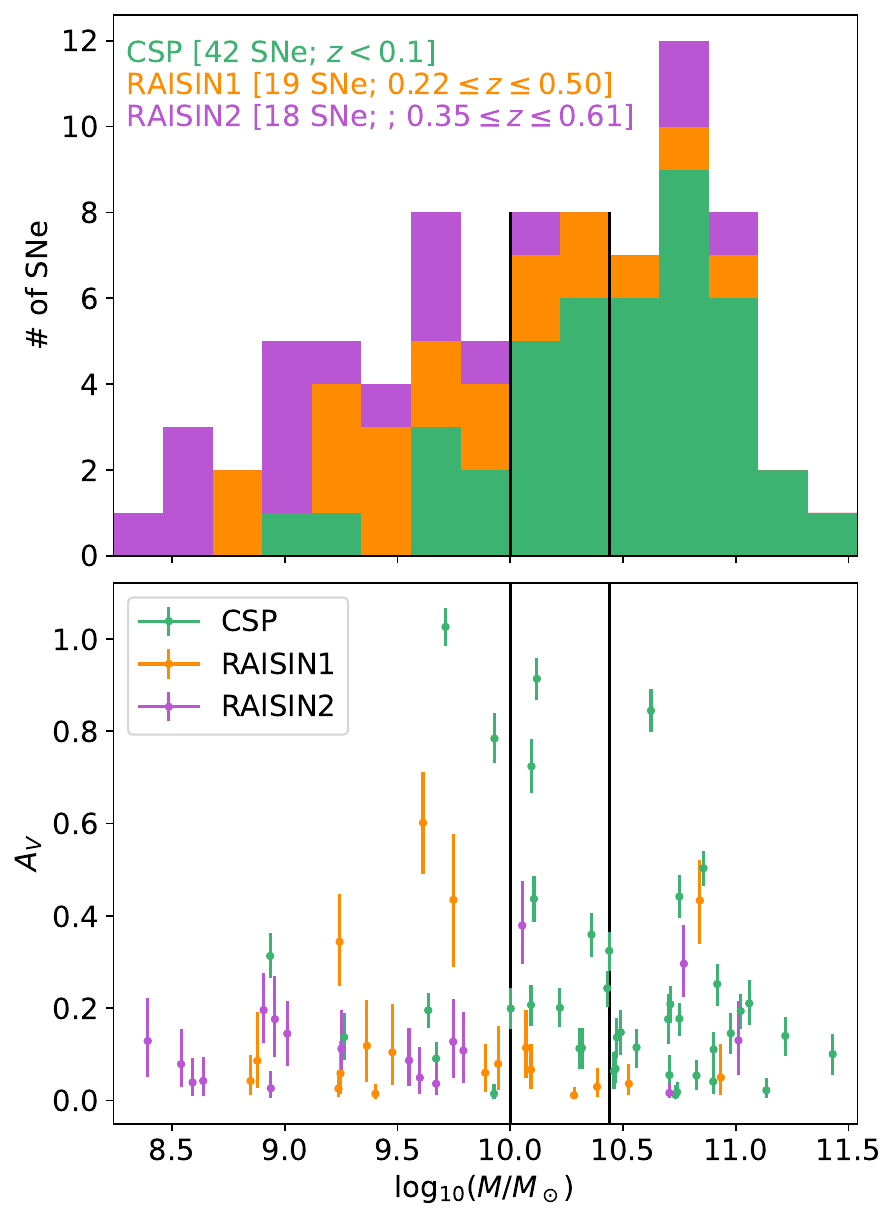}
    \caption{Estimated dust extinction ($A_V$) and host galaxy stellar mass ($\log_{10}(M/\mathrm{M}_\odot)$) for the CSP and RAISIN samples. Vertical lines correspond to host galaxy stellar masses of $10^{10}$ and $10^{10.44}~\mathrm{M}_\odot$. (top panel) Stacked histogram of $\log_{10}(M/\mathrm{M}_\odot$). (bottom panel) Estimated $A_V$ vs.\ $\log_{10}(M/\mathrm{M}_\odot$).}
    \label{fig:AV_vs_mass}
\end{figure}

Figure \ref{fig:AV_vs_mass} shows the $A_V$\footnote{Estimated from the inference performed in Section \ref{sec:global_z_split}, Fig.~\ref{fig:raisin_1+2_global}. The plotted points correspond to posterior medians and 68 per cent credible intervals. Estimates of $A_V$ are fairly insensitive to the sample division and exact treatment of $R_V$, so a similar distribution would be obtained from any of the analyses carried out in this paper.} and host galaxy stellar mass distribution for the CSP, RAISIN1, and RAISIN2 samples. This depicts graphically the mass distributions summarised in the paragraph above, and gives a sense of the number of SNe with moderate-to-high $A_V$ in each mass/redshift bin. The SNe with higher $A_V$ will have more influence on the $R_V$ inference, so the way these are distributed across surveys and host mass will inform us which SNe are driving the results in later sections.

\begin{figure}
    \centering
    \includegraphics[width=\linewidth]{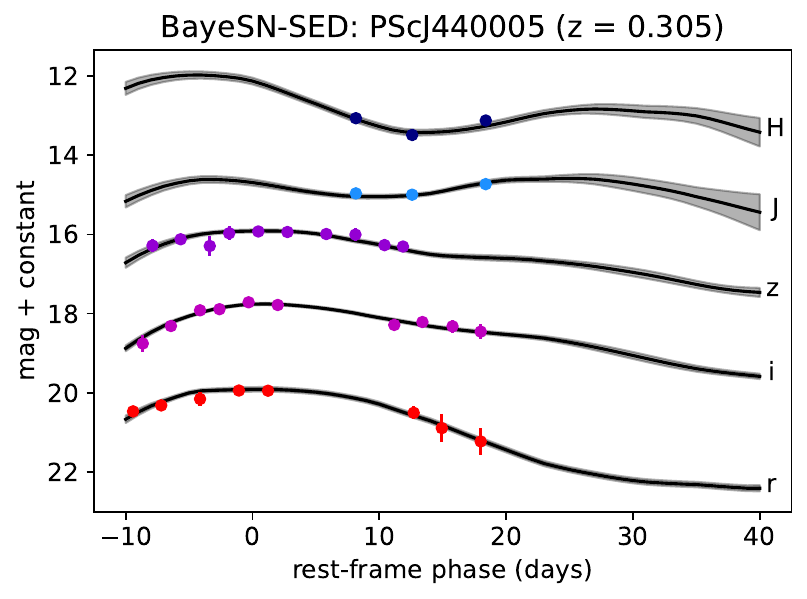}
    \includegraphics[width=\linewidth]{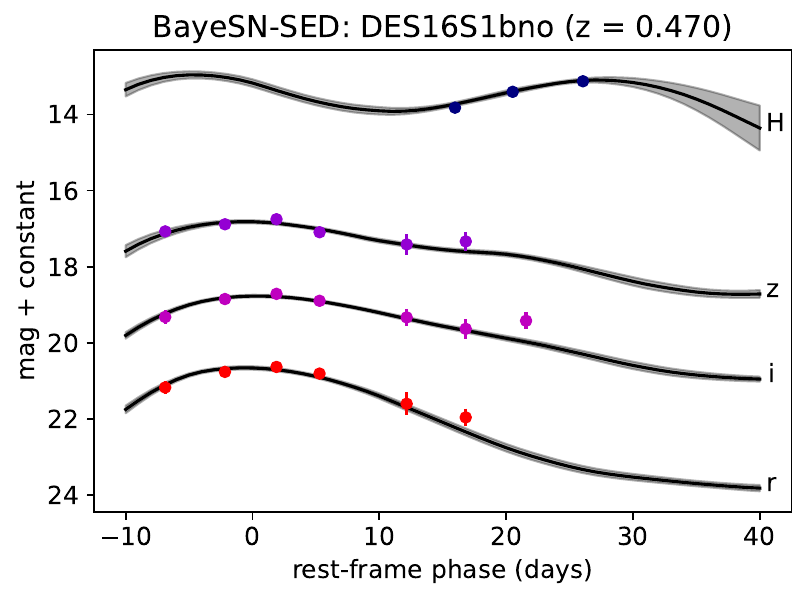}
    \caption{Examples of SN Ia light curve data from the RAISIN1 (upper panel) and RAISIN2 (lower panel) surveys. The RAISIN1 optical ($riz$) data are from the Pan-STARRS Medium Deep Survey, whilst for RAISIN2 these data are from the Dark Energy Survey. The $J$- and $H$-band data are from the \textit{Hubble Space Telescope} $F125W$ and $F160W$ filters. Also shown are fits to the data using the \citetalias{mandel22} \textsc{BayeSN} model. The vertical offsets applied for plotting are $(-2, -4, -6, -8, -11)$ for $(r,i,z,J,H)$.}
    \label{fig:light_curves}
\end{figure}

A full description of the photometric data can be found in \citet[\S2]{jones22}, with details regarding sample selection in \citet[\S3.4]{jones22}. Two example light curves from RAISIN are shown in Figure \ref{fig:light_curves}, together with fits made using the \textsc{BayeSN} model.

Given the oppositely skewed mass distributions of the CSP and RAISIN samples, one\footnote{We thank the referee for this suggestion.} can ask if there is a more optimal split that minimises the correlation between mass and redshift. This can be framed as finding the mass split that simultaneously maximises the fraction of the CSP sample in the low mass bin, and the fraction of the RAISIN sample in the high mass bin. Since these two objectives are in conflict, there is no perfect solution that brings them both close to the ideal value of $0.5$. One can then consider the set of Pareto-optimal mass splits, where a shift to lower or higher masses reduces at least one of the two objectives (for a review see \citealp{hwang79, miettinen98}). Neither $10^{10}$ nor $10^{10.44}~\mathrm{M}_\odot$ is a Pareto-optimal choice. However $10^{10}~\mathrm{M}_\odot$ is very close -- moving the step to the left or right by one SN is the only lossless improvement that can be made. To choose a preferred mass from among the Pareto-optimal set, we can solve a combined optimization problem,
\begin{equation}
    M^* = \argmin_M\left[(N^\text{CSP}_{<M}/42 - 0.5)^2 + (N^\text{RAISIN}_{\geq M}/37 - 0.5)^2\right],
\end{equation}
where the loss function is the sum of squared distances between the two objectives and their ideal value of $0.5$ \citep{yu73}. Here we use $N_C^S$ to denote the number of SNe in survey $S$ whose hosts satisfy criterion $C$. For our data, we find the optimal value to be in the range $10.360<\log_{10}(M^*/\mathrm{M}_\odot)< 10.385$. For a mass split in this range, $26/42$ CSP SNe would fall into the upper mass bin, and $7/37$ RAISIN SNe would fall into the upper mass bin. Since this is a very marginal improvement on the $10^{10.44}~\mathrm{M}_\odot$ split, we do not employ this optimized split in any of our subsequent analyses.

\section{Model}
\label{sec:model}
We perform our analysis using the \textsc{BayeSN} hierarchical Bayesian model for the spectral energy distributions (SEDs) of SNe Ia \citep{mandel22, grayling24}. This framework models variations in the time-dependent SN Ia SED in terms of distinct intrinsic and dust components, as summarised below. We use the \citetalias{mandel22} version of the model, presented in \citet{mandel22}, and trained on low-redshift $BVRIYJH$ light curves compiled by \citet{avelino19} from the CfA/CfAIR Supernova Programs \citep{hicken09, hicken12, woodvasey08, friedman15}, Carnegie Supernova Project \citep{krisciunas17}, and elsewhere in the literature \citep[see references in][]{avelino19, mandel22}. This is the same version of the model as was used in the previous analyses by \citet{jones22}, \citet{thorp22}, and \citet{dhawan22}. The model covers a rest-frame wavelength range of $\sim3500$--18500~\AA.

We adopt the same hierarchical Bayesian fitting procedure as \citet[\S3.2~(ii)]{thorp22}, whereby all supernovae in the sample (or a subsample thereof) are fitted simultaneously with the \citetalias{mandel22} \textsc{BayeSN} model. For each supernova, $s$, we fit for a set of latent parameters, $\bm{\phi}_s$. These include the distance modulus, $\mu_s$; host galaxy dust extinction, $A_V^s$; the light curve shape, $\theta_1^s$; a vector of residual intrinsic SED perturbations, $\bm{e}_s$; and a grey (i.e. constant in time and wavelength) luminosity offset, $\delta M_s$. Depending on our model configuration, we may also fit for individual values of the total-to-selective extinction ratio, $R_V^s$, along the line of sight in each supernova's host galaxy\footnote{We assume a \citet{fitzpatrick99} extinction law. For a discussion of this vs.\ \citet{cardelli89} or \citet{odonnell94}, see \citet{burns14} or \citet{thorp22}.}. At the population level, we fit for the population mean dust extinction, $\tau_A$, that parameterises an exponential population distribution. We also fit for either a single common $R_V$, or, in the case of a $R_V^s$ population distribution, its mean, $\mu_R$, and standard deviation, $\sigma_R$.  In the latter case, we sample the joint posterior distribution,
\begin{multline}
    P(\{\mu_s, \bm{\phi}_s\}, \mu_R, \sigma_R, \tau_A| \{\bm{\hat{f}}_s, z_s\}, \bm{\hat{W}}_0, \bm{\hat{W}}_1, \bm{\hat{\Sigma}}_\epsilon, \hat{\sigma}_0) \\\propto \bigg[\prod_s P(\bm{\hat{f}}_s | \mu_s, A_V^s, R_V^s, \theta_1^s, \delta_M^s, \bm{e}_s, \bm{\hat{W}}_0, \bm{\hat{W}}_1) \\ \times P(\mu_s | z_s) \times P(\theta_1^s) \times P(A_V^s|\tau_A)\times P(R_V^s|\mu_R,\sigma_R) \\ \times P(\bm{e}_s|\bm{\hat{\Sigma}}_\epsilon)\times P(\delta M_s|\hat{\sigma}_0)\bigg] \times P(\tau_A) \times P(\mu_R) \times P(\sigma_R),
    \label{eq:jointpost}
\end{multline}
following \citet[eq.~9]{thorp22}. Here, $\bm{\hat{W}}_0$, $\bm{\hat{W}}_1$, $\bm{\hat{\Sigma}}_\epsilon$, and $\hat{\sigma}_0$ are hyperparameter estimates obtained from model training. Respectively, these are the population mean SED function, a functional principal component capturing the primary mode of SED shape variation, a covariance matrix of intrinsic residual perturbations, and the level of grey residual brightness scatter. The first term inside the product is our flux data likelihood (constructed as per \citealp{mandel22}). The remaining terms in the product are our are our priors on the latent parameters:
\begin{align}
    P(\mu_s|\,z_s) &= N(\mu_s|\, \hat{\mu}_{\text{ext},s}, \hat{\sigma}_{\text{ext},s}^2), \label{eq:P_mu}\\
    P(A_V^s|\tau_A) &= \exponential(A_V^s|\tau_A), \label{eq:P_AV}\\
    P(R_V^s|\,\mu_R,\sigma_R) &= N(R_V^s| \mu_R, \sigma_R^2) \text{ for } R_V^s\geq0.5,\\
    P(\theta_1^s) &= N(\theta_1^s|\,0, 1^2),\label{eq:P_theta}\\
    P(\bm{e}_s|\bm{\hat{\Sigma}}_\epsilon) &= N(\bm{e}_s|\, \bm{0}, \bm{\hat{\Sigma}}_\epsilon),\\
    P(\delta M_s|\,\hat{\sigma}_0) &= N(\delta M_s|0, \hat{\sigma}_0^2).
\end{align}
Here, we set our distance constraint, $P(\mu_s | z_s)$, based on an external redshift-based distance estimate, $\hat{\mu}_{\text{ext},s}$, and its uncertainty, $\hat{\sigma}_{\text{ext},s}$. Our method for estimating these is given in Section \ref{sec:data}. For our hyperpriors, we adopt
\begin{align}
    P(\tau_A) &= \halfcauchy(\tau_A|0,1),\label{eq:P_tau}\\
    P(\mu_R) &= U(\mu_R|1,5),\\
    P(\sigma_R) &= \text{Half-}N(\sigma_R|0,2^2)\label{eq:P_sigmaR},
\end{align}
following \citet{mandel22}, \citet{thorp21}, and \citet{thorp22}.

When fitting for a single common $R_V$ within a sample or subsample, we sample the joint posterior distribution given by
\begin{multline}
    P(\{\mu_s, \bm{\phi}_s\}, R_V, \tau_A| \{\bm{\hat{f}}_s, z_s\}, \bm{\hat{W}}_0, \bm{\hat{W}}_1, \bm{\hat{\Sigma}}_\epsilon, \hat{\sigma}_0) \\\propto \bigg[\prod_s P(\bm{\hat{f}}_s | \mu_s, A_V^s, R_V, \theta_1^s, \delta_M^s, \bm{e}_s, \bm{\hat{W}}_0, \bm{\hat{W}}_1) \times P(\mu_s | z_s)) \times P(\theta_1^s) \\ \times P(A_V^s|\tau_A)  \times P(\bm{e}_s|\bm{\hat{\Sigma}}_\epsilon)\times P(\delta M_s|\hat{\sigma}_0)\bigg] \times P(\tau_A) \times P(R_V).
    \label{eq:jointpost_global}
\end{multline}
The priors and hyperpriors here are as given in Equations \ref{eq:P_mu}, \ref{eq:P_AV}, \ref{eq:P_theta}--\ref{eq:P_tau}, with our hyperprior on the common $R_V$ being given by
\begin{equation}
    P(R_V) = U(R_V|1,6).
\end{equation}

Complete details of the \textsc{BayeSN} model can be found in \citet{mandel22}, with specific discussion regarding its application to dust being included in \citet{thorp21}, \citet{thorp22}, and \citet{grayling24}. As described therein, to fit these statistical models to the data, we use Hamiltonian Monte Carlo (within Stan; \citealp{carpenter17}) to sample the posterior distributions above and use standard diagnostics \citep{gelman92, betancourt14, betancourt16, vehtari21} to assess the chains.

\section{Results}
\label{sec:results}

\subsection{Common Dust Law Inference}
\label{sec:global_results}
In this Section, we present results from estimating the best fitting common $R_V$ value for different subsamples of our data. Whilst assuming a common $R_V$ in each subsample is a strong assumption, it allows us to obtain tighter constraints than the more flexible model where an $R_V$ population distribution of non-zero width is permitted (Section \ref{sec:pop_results} presents our results under this more relaxed assumption). The assumption that SNe in a particular redshift range might be well modelled by a single common $R_V$ is an approximation, but it can still provide important information about the typical nature of dust in the subsamples analysed.

In Section \ref{sec:survey_global_z_split}, we present $R_V$ inferences for CSP, RAISIN 1, and RAISIN2 separately. We will then proceed in Section \ref{sec:global_z_split} to present our inferences of $R_V$ for CSP ($z<0.1$) and the joint RAISIN1+2 ($0.22\leq z\le0.61$) sample. In Section \ref{sec:global_z+m_split}, we will further split these two subsamples by host galaxy stellar mass -- testing if there is mass dependence of $R_V$ in either redshift bin, and if this may differ as a function of redshift.

\subsubsection{Split by Survey}
\label{sec:survey_global_z_split}
\begin{figure}
    \centering\includegraphics[width=\linewidth]{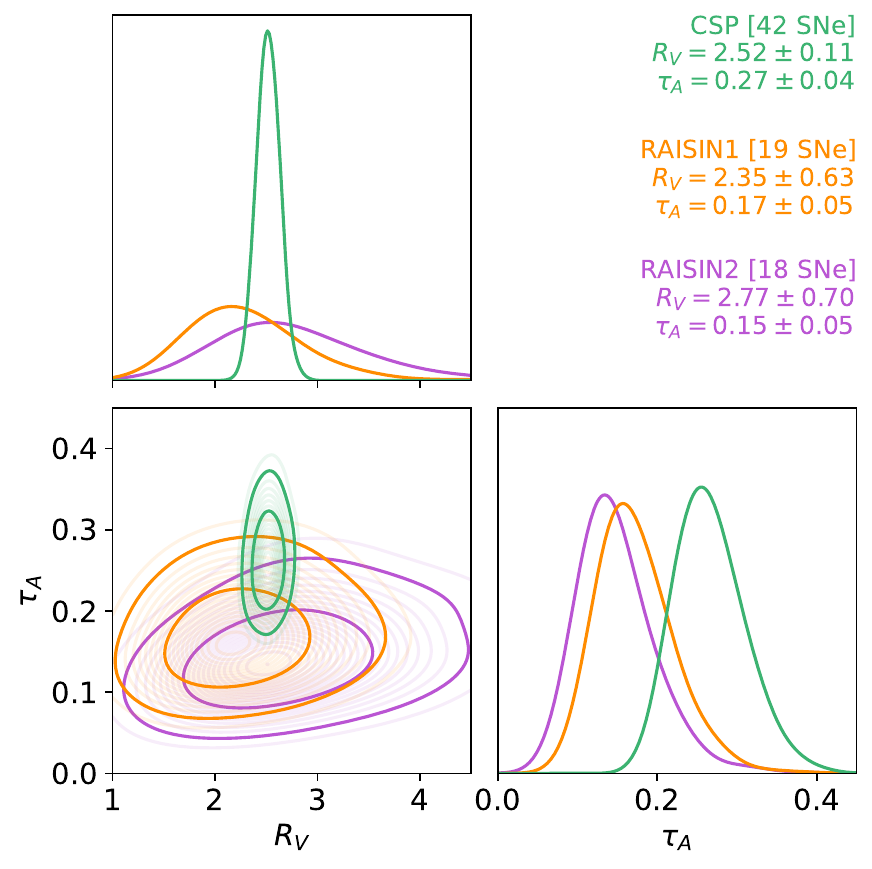}
    \caption{Posterior distributions of the best fitting dust law $R_V$, and population mean extinction, $\tau_A$, for the three subsamples of our data. Green contours show the CSP sample ($z<0.1$), whilst orange and purple contours show RAISIN1 ($0.22\leq z\le0.50$) and RAISIN2 ($0.35\leq z\leq0.61$), respectively. Parameter summaries are posterior mean $\pm$ standard deviation.}
    \label{fig:raisin_1+2_global}
\end{figure}

In the first instance, we perform our inference within three subsamples: CSP ($z<0.1$), RAISIN1 ($0.22\leq z \leq 0.50$), and RAISIN2 ($0.35\leq z \leq 0.61$). Figure \ref{fig:raisin_1+2_global} shows our posterior distributions of $R_V$, and the mean dust extinction, $\tau_A$, for these three samples. These results show strong consistency between the posterior distributions for the two RAISIN samples (purple and orange contours in Fig.~\ref{fig:raisin_1+2_global}), so we will combine these in subsequent analyses into a single higher-$z$ RAISIN1+2 sample, which will have greater constraining power.

\subsubsection{Split by Redshift}
\label{sec:global_z_split}
\begin{figure}
    \centering\includegraphics[width=\linewidth]{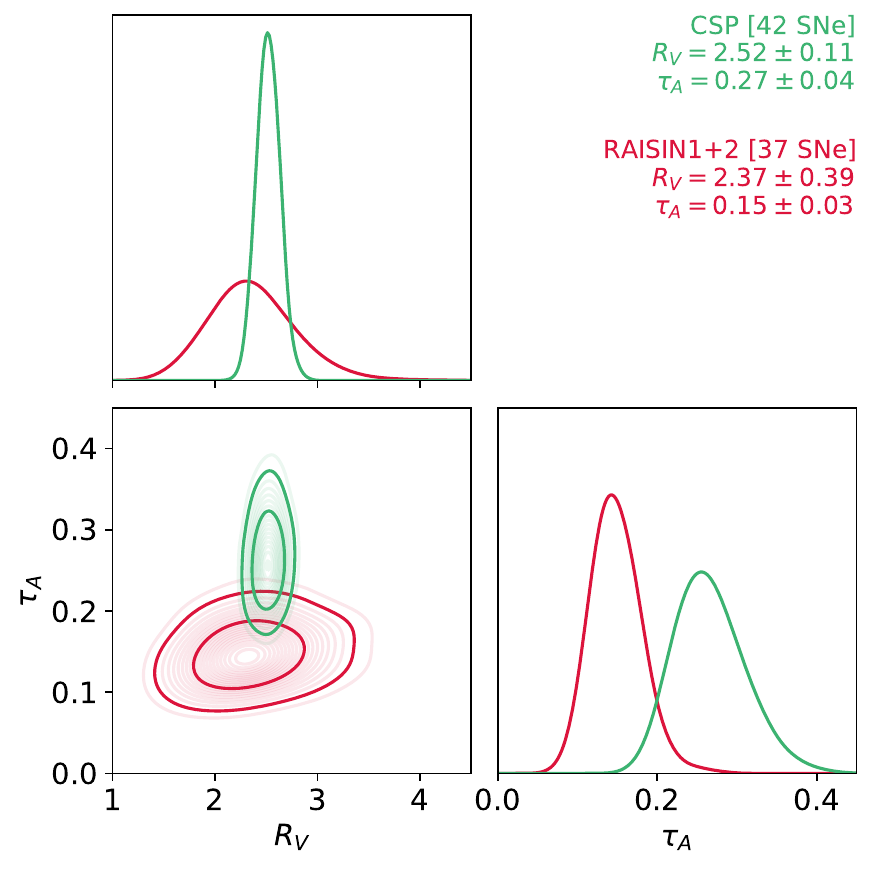}
    \caption{Same as Figure \ref{fig:raisin_1+2_global}, but showing results where the two RAISIN subsamples were merged and assumed to have a common $R_V$ and extinction distribution. Red contours show the inference for this combined RAISIN1+2 sample. Green contours show CSP, as in Figure \ref{fig:raisin_1+2_global}.}
    \label{fig:raisin_global}
\end{figure}

Figure \ref{fig:raisin_global} shows our posterior distributions of $R_V$ and $\tau_A$ for the CSP sample (indicated by the green contours, which are the same as in Fig.~\ref{fig:raisin_1+2_global}), and for the combined RAISIN1+2 sample (red contours). For CSP, we estimate $R_V=2.52\pm0.11$, whilst for RAISIN1+2, we estimate $R_V=2.37\pm0.39$, consistent with CSP to within the uncertainties. The estimate we obtain here for CSP is highly consistent with the population mean $R_V$ ($\mu_R=2.50\pm0.12$) estimated by \citet{thorp22} using a slightly larger sample of 75 CSP SNe Ia with apparent $B-V\leq 0.3$. For CSP, we estimate $\tau_A=0.27\pm0.04$~mag, somewhat higher that the estimate of $\tau_A=0.15\pm0.03$ that we obtain for RAISIN1+2. This is unsurprising, as it is likely that the RAISIN sample will be biased towards low-$A_V$ SNe due to the difficulty of detecting these objects at high redshift (see also the discussion of similar results in the optical analysis of \citealp{grayling24}). We defer a Bayesian treatment of this selection effect to future work.

\begin{figure}
    \centering
    \includegraphics[width=\linewidth]{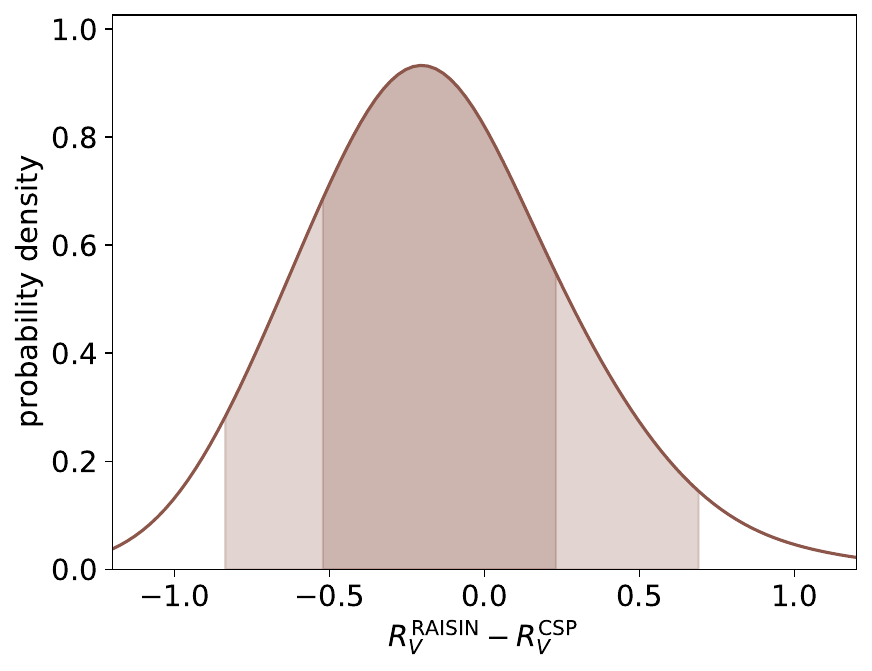}
    \caption{Posterior probability distribution of the $R_V$ difference between the high-$z$ RAISIN sample, and low-$z$ CSP sample. Shaded regions show 68 and 95 per cent credible intervals.}
    \label{fig:delta_RV_z_split}
\end{figure}

From the posterior samples shown in Figure \ref{fig:raisin_global}, we can derive a posterior distribution of the high- vs.\ low-$z$ (i.e.\ RAISIN vs.\ CSP) $R_V$ difference ($\Delta R_V^z = R_V^\text{RAISIN} - R_V^\text{CSP}$). Figure \ref{fig:delta_RV_z_split} shows this. From this, we can estimate that $-0.84<\Delta R_V^z<0.71$, with 95 per cent posterior probability. This is fully consistent with a picture in which there is no evolution in the typical $R_V$ as a function of redshift, but does not rule out a shift in either direction.

\subsubsection{Split by Redshift and Host Galaxy Mass}
\label{sec:global_z+m_split}
In this section, we perform our inference with the CSP and RAISIN samples each being split by host galaxy stellar mass, giving four subsamples in total. We follow \citet{jones22} in choosing the sample split points to test, considering splits at $10^{10}~\mathrm{M}_\odot$, and at $10^{10.44}~\mathrm{M}_\odot$. The latter corresponds to a best-fitting NIR mass step location identified by \citet{ponder21} using the Akaike Information Criterion.

\begin{figure*}
    \centering
    \includegraphics[width=0.5\linewidth]{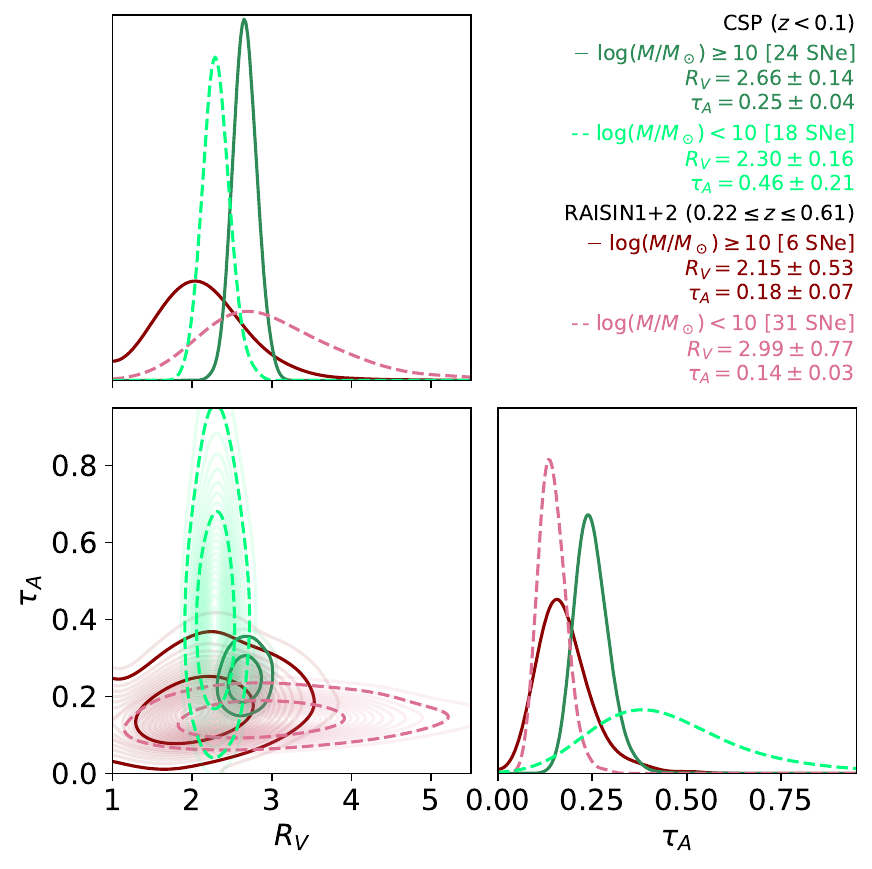}%
    \includegraphics[width=0.5\linewidth]{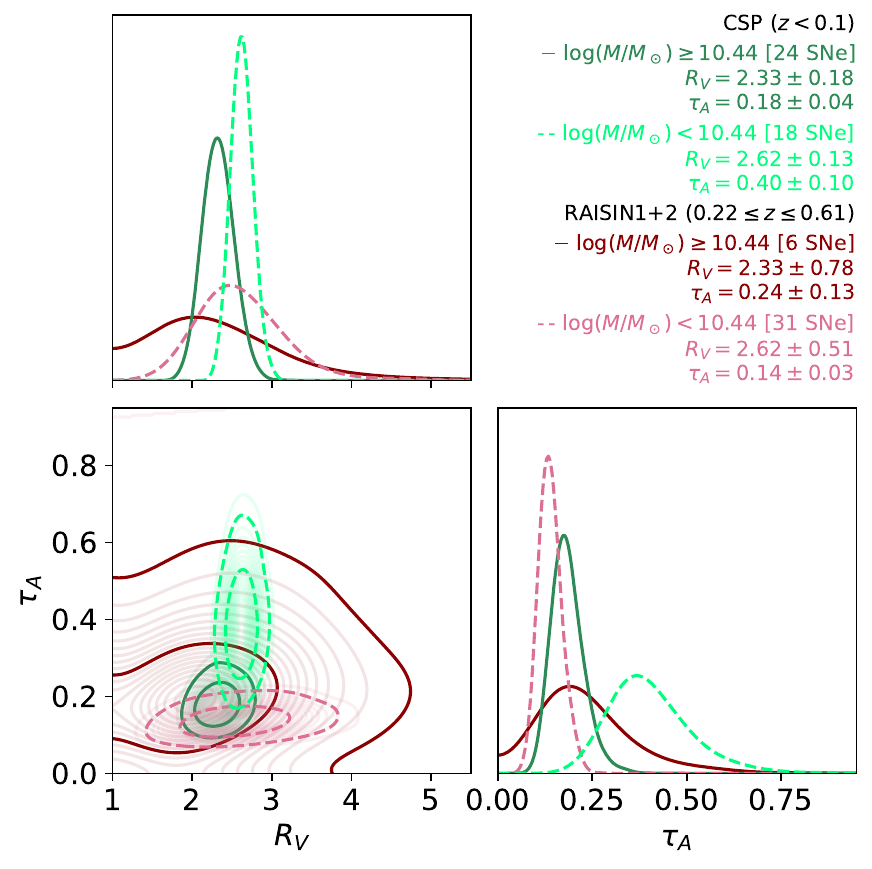}
    \caption{Same as Figure \ref{fig:raisin_global}, but with the low-redshift CSP sample ($z<0.1$) and higher-redshift RAISIN sample ($0.22\leq z\leq 0.61$) each subdivided by host galaxy stellar mass. Green contours show CSP, whilst red/pink contours show RAISIN. Darker solid contours show higher host galaxy masses, whilst lighter dashed contours show lower host masses. We use the two sample split points used by \citet{jones22} as their default choice and systematic test. (left panels) Inference when the sample is split at the default host galaxy stellar mass of $10^{10}~\mathrm{M}_\odot$. (right panels) Split at $10^{10.44}~\mathrm{M}_\odot$, the best fitting NIR mass step location found by \citet{ponder21}.}
    \label{fig:raisin_global_mass_split}
\end{figure*}

Figure \ref{fig:raisin_global_mass_split} shows our posterior distributions for $R_V$ and $\tau_A$ in the low- and high-host mass RAISIN and CSP samples. The left hand panels show the results for the $10^{10}~\mathrm{M}_\odot$ mass split, with the right showing the split at $10^{10.44}~\mathrm{M}_\odot$. In terms of $R_V$, the picture is generally one of consistency to within the uncertainties. For CSP, we estimate $R_V=2.30\pm0.16$ for host galaxies less massive than $10^{10}~\mathrm{M}_\odot$, and $2.66\pm0.14$ for more massive hosts. In the higher-redshift bin, the estimates of $R_V$ are more uncertain. Within these uncertainties, the $R_V$ estimates for the two mass bins are consistent with one another, and with their low-redshift counterparts.

Following on from Fig.~\ref{fig:delta_RV_z_split} in Section \ref{sec:global_z_split}, we can use our posterior samples to derive probability distributions of the change in $R_V$ with either redshift (in the two mass bins), or mass (in the two redshift bins). This allows us to investigate two related questions:
\begin{enumerate}
    \item Is the potential evolution of $R_V$ with redshift ($\Delta R_V^z = R_V^\text{RAISIN}-R_V^\text{CSP}$) different for low- and high-mass host galaxies?
    \item Are the potential correlations of $R_V$ with mass ($\Delta R_V^M = R_V^{\text{high-}M}-R_V^{\text{low-}M}$) different at low- and high-redshift?
\end{enumerate}

\begin{figure*}
    \centering
    \includegraphics[width=0.5\linewidth]{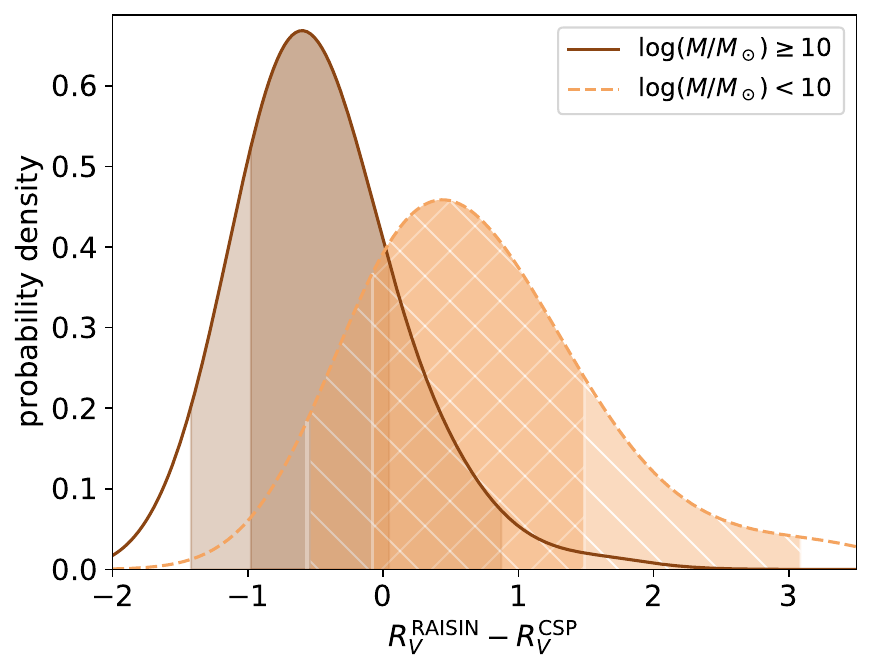}%
    \includegraphics[width=0.5\linewidth]{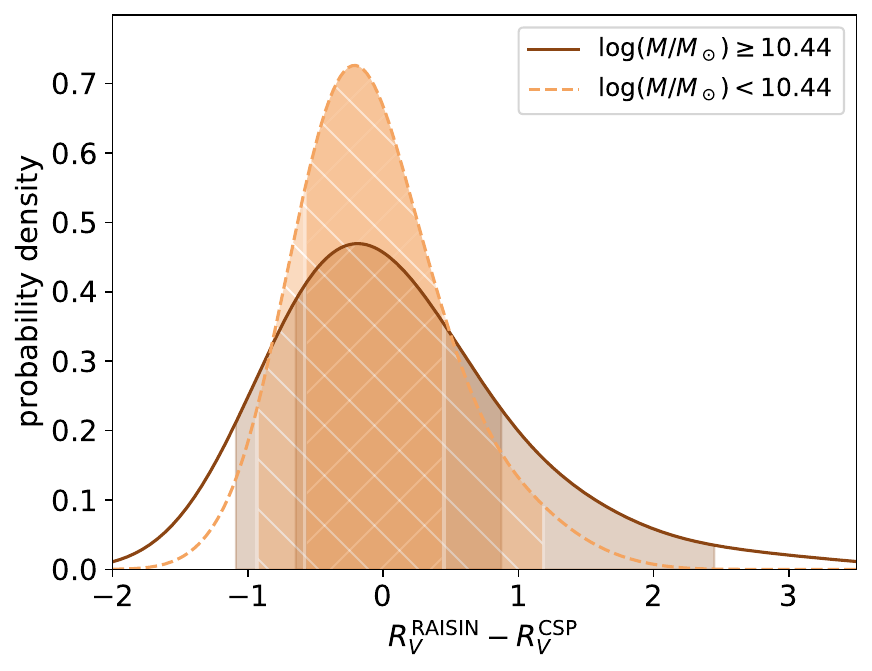}
    \caption{Same as Figure \ref{fig:delta_RV_z_split}, but with the sample split into two bins of high (solid dark brown curves) and low (dashed light brown curves) host galaxy mass. (left panel) Sample divided at $10^{10}~\mathrm{M}_\odot$. (right panel) Sample divided at $10^{10.44}~\mathrm{M}_\odot$.}
    \label{fig:delta_RV_z+mass_split}
\end{figure*}

We will tackle the first question first, and will estimate the posterior probability distributions of $\Delta R_V^z = R_V^\text{RAISIN}-R_V^\text{CSP}$ separately for high- and low-mass host galaxies. Figure \ref{fig:delta_RV_z+mass_split} shows the resulting posteriors for the $10^{10}~\mathrm{M}_\odot$ and $10^{10.44}~\mathrm{M}_\odot$ mass splits. For both mass splits, the picture is fairly uncertain, but there is not significant evidence that the high- vs.\ low-$z$ $\Delta R_V^z$ differs for low- and high-mass hosts. For the split at $10^{10}~\mathrm{M}_\odot$, the $\Delta R_V^z$ distribution for high-mass hosts leans towards negative values (i.e.\ lower $R_V$ at higher $z$), whereas the $\Delta R_V^z$ distribution for low-mass hosts leans towards positive values (i.e.\ higher $R_V$ at higher $z$). Zero is within the 68 per cent credible interval for both mass bins, however, and both bins have significant posterior density at both positive and negative values of $\Delta R_V^z$. For the split at $10^{10.44}~\mathrm{M}_\odot$ (left panel of Fig.~\ref{fig:delta_RV_z+mass_split}), the picture is similar, although the $\Delta R_V^z$ posteriors in the two mass bins are more similar in this case, both having their modes very close to $\Delta R_V^z=0$, and having slightly longer tails towards positive $\Delta R_V^z$. 

\begin{figure*}
    \centering
    \includegraphics[width=0.5\linewidth]{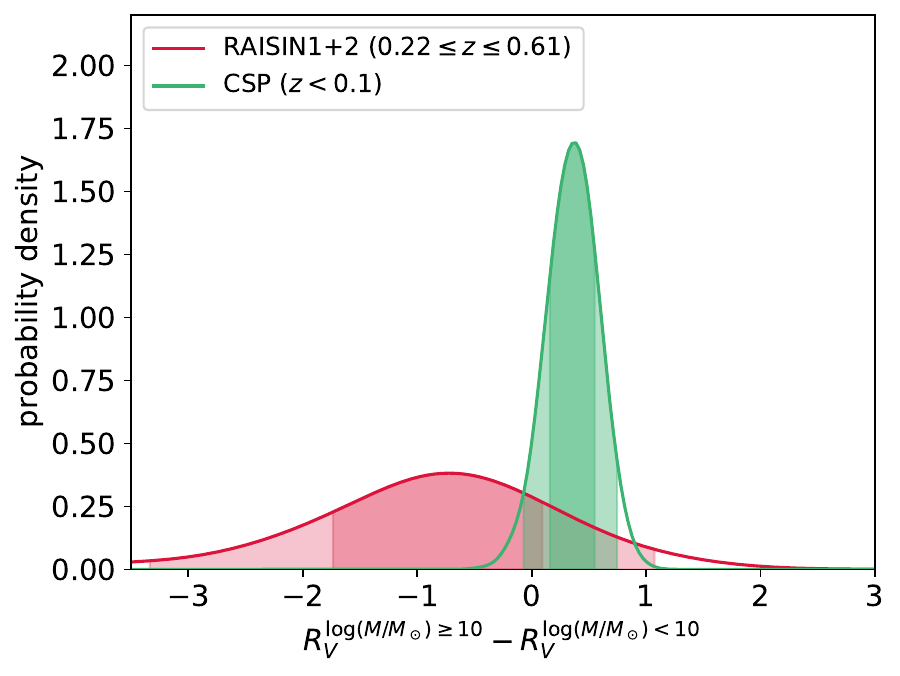}%
    \includegraphics[width=0.5\linewidth]{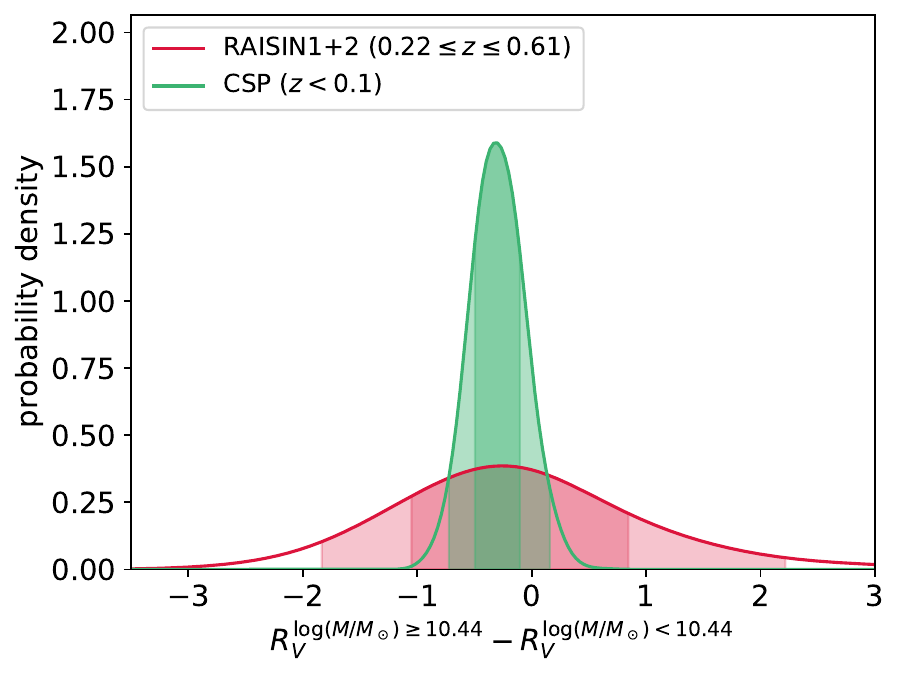}
    \caption{Same as Figure \ref{fig:delta_RV_z+mass_split}, but showing high- vs.\ low-mass $\Delta R_V^M$ for the two redshift bins (rather than high- vs.\ low-$z$ $\Delta R_V^z$ for the two mass bins). (left panel) Sample divided at $10^{10}~\mathrm{M}_\odot$. (right panel) Sample divided at $10^{10.44}~\mathrm{M}_\odot$.}
    \label{fig:delta_RV_mass+z_split}
\end{figure*}

To tackle the second question, we will estimate the posterior distributions of $\Delta R_V^M = R_V^{\text{high-}M}-R_V^{\text{low-}M}$ separately for the RAISIN (high-$z$) and CSP (low-$z$) samples. Figure \ref{fig:delta_RV_mass+z_split} shows this for the $10^{10}~\mathrm{M}_\odot$ and $10^{10.44}~\mathrm{M}_\odot$ mass splits. For both mass splits, the resulting posterior distribution is much tighter for the CSP sample than RAISIN. The results are consistent for the two mass splits, and do not provide statistically significant evidence for a changing $\Delta R_V^M$ between low- and high-redshift. For the CSP sample, the $\Delta R_V^M$ posterior has more probability towards positive values (i.e.\ higher $R_V$ in higher mass galaxies) for the $10^{10}~\mathrm{M}_\odot$ mass split, but more towards negative values for the $10^{10.44}~\mathrm{M}_\odot$ mass split. The latter is more in line with our previous results at low-$z$ (see \citealp{thorp21,thorp22}), although both there and here the estimated $\Delta R_V^M$ (or $\Delta \mu_R^M$ in \citealp{thorp22}) has not been significantly different from zero. Indeed, for both mass splits considered here, $\Delta R_V^M=0$ is within the 95 per cent posterior credible intervals for CSP. For RAISIN, the posteriors show a mild tendency towards negative $\Delta R_V^M$ (i.e.\ lower $R_V$ for higher mass host galaxies) for both choices of mass split, although $\Delta R_V^M=0$ is within the 68 per cent credible interval in both cases. The results for the $10^{10}~\mathrm{M}_\odot$ mass split would be consistent with a larger potential redshift evolution of $\Delta R_V^M$ than the results for $10^{10.44}~\mathrm{M}_\odot$, although due to the large uncertainties in both cases, there is no statistically significant evidence for such a redshift evolution in either case. 

\subsection{Dust Law Population Distribution Inference}
\label{sec:pop_results}
In this Section, we relax the assumption that each SN Ia subsample is consistent with a single $R_V$ common to that subsample. Instead, we assume a Gaussian population distribution of $R_V$, with the $R_V^s$ for a supernova, $s$, being modelled as a draw from this population distribution, like so: $R_V^s\sim N(\mu_R, \sigma_R^2)$. When we perform our inference, we sample from the joint posterior distribution given by Equation \ref{eq:jointpost}, conditional on the data for a given supernova subsample. We marginalize over all supernova-level parameters, including the set of individual $R_V^s$ values, to obtain a posterior distribution over the $R_V$ population mean, $\mu_R$, and standard deviation, $\sigma_R$.

In Section \ref{sec:pop_z_split}, we present our results for the low-$z$ CSP data, and higher-$z$ RAISIN data. In Section \ref{sec:pop_m_split}, we present results obtained by combining the CSP and RAISIN samples, and then splitting this combined sample by host galaxy stellar mass.

\subsubsection{Split by Redshift}
\label{sec:pop_z_split}

\begin{figure}
    \centering
    \includegraphics[width=\linewidth]{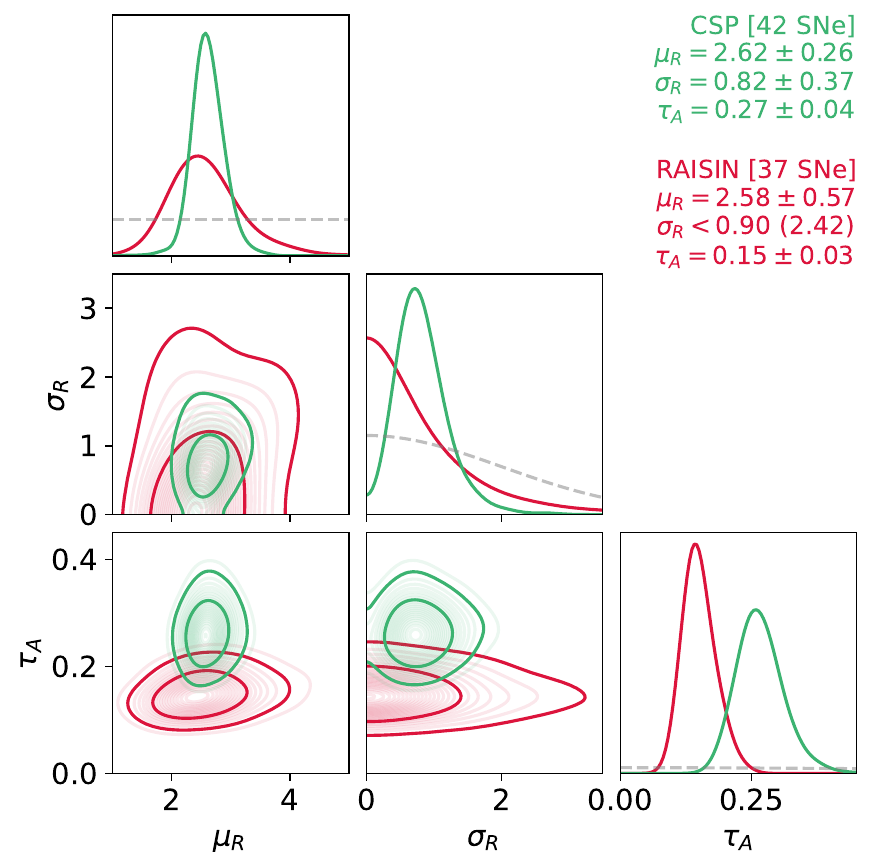}
    \caption{Posterior distributions of the host galaxy dust population distribution parameters for the CSP ($z<0.1$; green contours) and RAISIN ($0.2\lesssim z \lesssim 0.6$; red countours) components of our sample. Parameters plotted are the mean ($\mu_R$) and standard deviation ($\sigma_R$) of the Gaussian $R_V$ population distribution, and the mean ($\tau_A$) of the exponential $A_V$ distribution. Parameter summaries are ether posterior mean $\pm$ standard deviation, or 68~(95)th posterior percentile. Dashed lines indicate the hyperpriors.}
    \label{fig:raisin_pop}
\end{figure}

As in Section \ref{sec:global_z_split}, we will begin by performing an analysis with the sample split by redshift. Figure \ref{fig:raisin_pop} shows our posterior distributions of the $R_V$ population mean ($\mu_R$) and standard deviation ($\sigma_R$), and $A_V$ population mean ($\tau_A$). Our estimates of $\tau_A$ for the two samples are identical to those obtained from the analysis in the previous sections (see Fig.~\ref{fig:raisin_global} and Section \ref{sec:global_z_split}). We have previously found that inference of the mean dust extinction, $\tau_A$, is very insensitive to the assumption of a single $R_V$ or population distribution thereof, so this result is unsurprising (see \citealp{thorp21}).

For CSP, we estimate an $R_V$ population mean of $\mu_R=2.62\pm0.26$, and a population standard deviation of $\sigma_R=0.82\pm0.37$. This population mean estimate is consistent with the common $R_V=2.52\pm0.11$ estimated in Section \ref{sec:global_z_split}. Additionally, these population mean and standard deviation estimates are consistent with the estimates obtained by \citet{thorp22} using a larger sample of 75 CSP SNe Ia with apparent $B-V\leq0.3$. For RAISIN, we estimate $\mu_R=2.58\pm0.57$, fully consistent with the CSP population mean estimate. Our posterior distribution for $\sigma_R$ peaks at zero, meaning we can use the posterior credible intervals to place upper limits on $\sigma_R$ in the RAISIN redshift range. We estimate that $\sigma_R<0.90$ with 68 per cent posterior probability, and $\sigma_R<2.42$ with 95 per cent posterior probability.

\begin{figure}
    \centering
    \includegraphics[width=\linewidth]{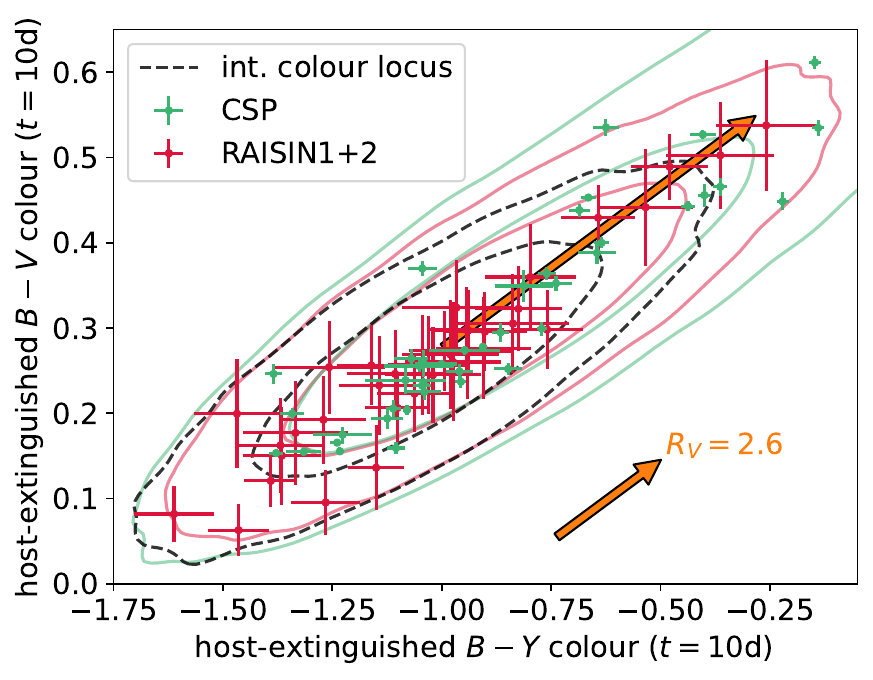}
    \caption{Estimated host-extinguished apparent $B-V$ vs.\ $B-Y$ colours for RAISIN (red points) and CSP (green points) at a rest-frame phase of 10~d. The red contours show the posterior predictive distribution (PPD) for RAISIN's host-extinguished colours, with the green showing the PPD for CSP's colours, from our dust law population inference (\S\ref{sec:pop_z_split}, Fig.\ \ref{fig:raisin_pop}). The black dashed contours show the intrinsic colour locus predicted by the \citetalias{mandel22} \textsc{BayeSN} model. The contours shown are 68 and 95 per cent credible intervals. The orange arrows show a reddening vector corresponding to $R_V=2.6$ (approximately the posterior mean of $\mu_R$ for CSP and RAISIN; c.f.\ Fig.\ \ref{fig:raisin_pop}). The longer arrow is has its base at the mean of the intrinsic colour locus and has a length corresponding to $A_V=0.75$. The shorter arrow has a length corresponding to $A_V=0.25$.}
    \label{fig:colours}
\end{figure}

Figure \ref{fig:colours} shows the estimated host-extinguished apparent $B-V$ and $B-Y$ colours for the RAISIN and CSP samples at a rest frame phase of 10~d. We chose this phase as a trade-off between being close to the data (the median phase of first observation for RAISIN is $\lesssim9.5$~d; \citealp{jones22}), but also being at a phase where intrinsic colour variation is relatively moderate (both due to correlation with light curve shape, and residual scatter on top of this; c.f.\ \citealp{mandel22} fig.\ 8 \& 9). The contours in our Figure \ref{fig:colours} show the posterior predictive distributions (PPD) for host-extinguished colours for the two redshift ranges, i.e.
\begin{multline}
    P[(B-V)_{t=10\,\mathrm{d}}^\text{ext}, (B-Y)_{t=10\,\mathrm{d}}^\text{ext}|\{\bm{\hat{f}}_s\}, \bm{\hat{W}}_0, \bm{\hat{W}}_1, \bm{\hat{\Sigma}}_\epsilon] 
    \\= \int P(\{\mu_s, \bm{\phi}_s\}, \mu_R, \sigma_R, \tau_A| \{\bm{\hat{f}}_s\}, \bm{\hat{W}}_0, \bm{\hat{W}}_1, \bm{\hat{\Sigma}}_\epsilon, \hat{\sigma}_0)
    \\ \times P[(B-V)_{t=10\,\mathrm{d}}^\text{ext}, (B-Y)_{t=10\,\mathrm{d}}^\text{ext}|\bm{\phi}]
    \times P(\bm{\phi}|\mu_R, \sigma_R, \tau_A, \bm{\hat{\Sigma}}_\epsilon)
    \\\times d\mu_s\, d\bm{\phi}_s\, d\mu_R\, d\sigma_R\, d\tau_A,
\end{multline} 
where
\begin{equation}
    P(\bm{\phi}|\mu_R, \sigma_R, \tau_A, \bm{\hat{\Sigma}}_\epsilon) = P(R_V|\mu_R, \sigma_R)P(A_V|\tau_A)P(\theta_1)P(\bm{e}|\bm{\hat{\Sigma}}_\epsilon).
\end{equation}
The PPDs are estimated by simulating supernovae from $P(\bm{\phi}|\mu_R, \sigma_R, \tau_A, \bm{\hat{\Sigma}}_\epsilon)$, conditional on the $\mu_R$, $\sigma_R$, and $\tau_A$ for each MCMC sample from the posteriors shown in Fig.\ \ref{fig:raisin_pop}, and computing their $B-V$ and $B-Y$ colours. Also shown in Fig.\ \ref{fig:colours} is the intrinsic colour locus of the \citetalias{mandel22} \textsc{BayeSN} model, determined by taking draws from the prior $P(\theta_1)\times P(\bm{e}|\bm{\hat{\Sigma}}_\epsilon)$ and estimating colours with $A_V=0$. The PPD corresponding to the low-$z$ CSP sample is more dispersed along the diagonal than for the higher-$z$ RAISIN sample, reflective of the posterior for CSP in Fig.\ \ref{fig:raisin_pop} being weighted towards a higher mean dust extinction, $\tau_A$. Although the constraints on the population standard deviation, $\sigma_R$, of $R_V$ for RAISIN are more uncertain than for CSP, the posterior for CSP is more concentrated towards non-zero values, with a posterior mean of 0.82. In the PPD, this is reflected as a broader distribution of colours perpendicular to the diagonal -- particularly towards redder end.

\begin{figure}
    \centering
    \includegraphics[width=\linewidth]{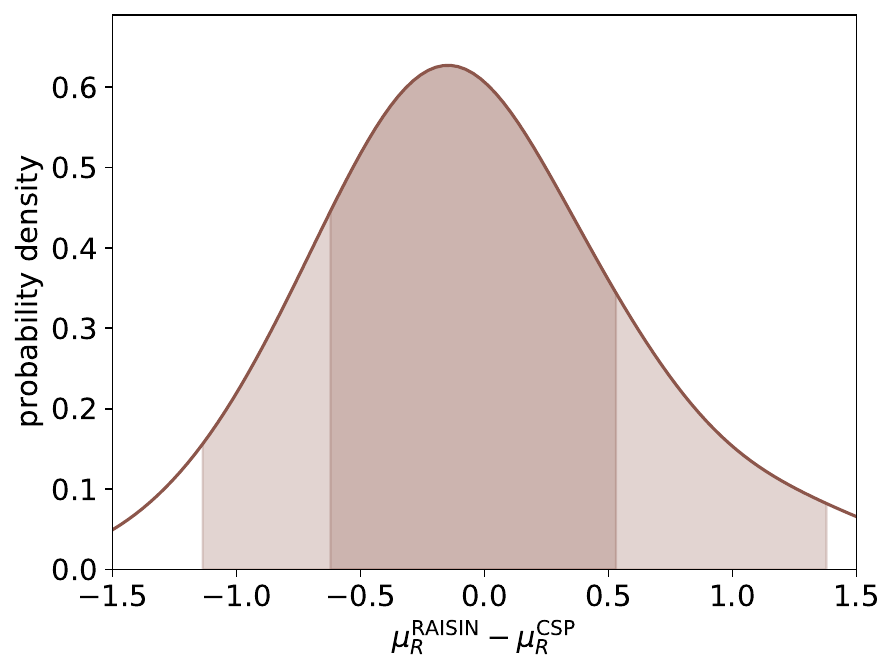}
    \caption{Posterior probability distribution of the estimated high- vs. low-$z$ difference in population mean $R_V$ ($\Delta\mu_R^z$).}
    \label{fig:delta_muR_z_split}
\end{figure}

Similarly to in Figure \ref{fig:delta_RV_z_split}, we can compute a derived posterior on the difference in $\mu_R$ ($\Delta \mu_R^z = \mu_R^\text{RAISIN}-\mu_R^\text{CSP}$) between the low-$z$ CSP and high-$z$ RAISIN samples. Figure \ref{fig:delta_muR_z_split} shows this. This posterior distribution is reasonably symmetric, so does not suggest a preference for either of the possible directions of $\mu_R$ shift. We can place limits on the possible redshift drift of $\mu_R$, estimating that $-0.63<\Delta \mu_R^z<0.55$ with 68 per cent posterior probability, and $-1.16<\Delta \mu_R^z<1.38$ with 95 per cent posterior probability.

\subsubsection{Split by Host Galaxy Mass}
\label{sec:pop_m_split}
Motivated by the consistency between the CSP and RAISIN samples seen in Section \ref{sec:pop_z_split}, in this section we combine the data from the two surveys, and then perform our inference with this joint sample split by host galaxy mass. When fitting the $R_V$ population distribution model (Eq.~\ref{eq:jointpost}), we lack the leverage to split the sample four ways, as we did in Section \ref{sec:global_z+m_split}. Nevertheless, we can still utilise this combined sample to investigate the possible dependence of the $R_V$ population distribution on host galaxy stellar mass.

\begin{figure*}
    \centering
    \includegraphics[width=0.5\linewidth]{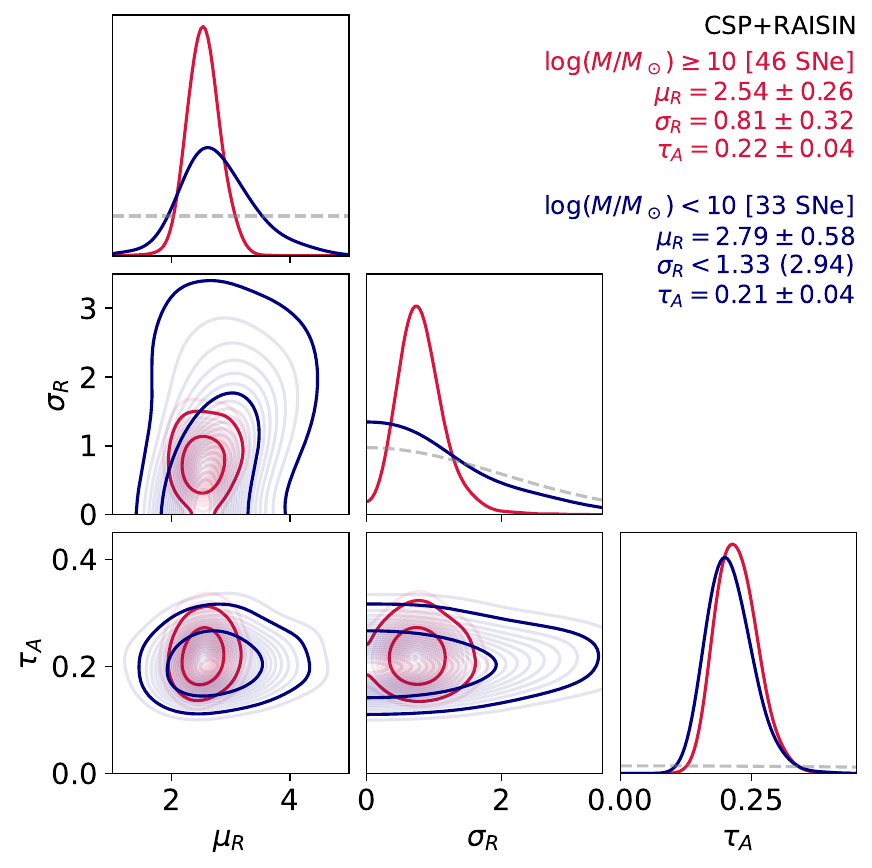}%
    \includegraphics[width=0.5\linewidth]{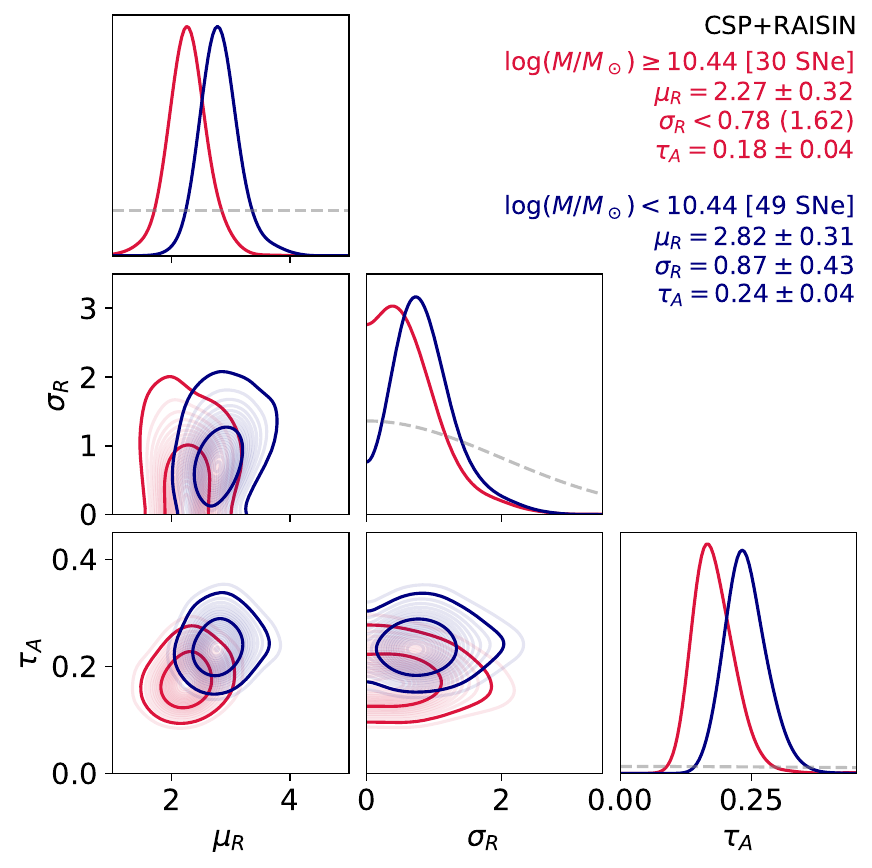}
    \caption{Posterior distributions of the host galaxy dust population distribution parameters of the combined CSP+RAISIN sample, split by host galaxy stellar mass. Red contours show the higher mass subsample, whilst blue show the lower mass subsample. Parameters plotted are the same as in Fig.~\ref{fig:raisin_pop}. Mass splits in the left- and right-hand corner plots are the same as in Figure \ref{fig:raisin_global_mass_split}.}
    \label{fig:raisin_pop_mass_split}
\end{figure*}

Figure \ref{fig:raisin_pop_mass_split} shows our posterior inference of $\mu_R$, $\sigma_R$, and $\tau_A$ for the mass split CSP+RAISIN sample. We test the same pair of split points ($10^{10}~\mathrm{M}_\odot$ and $10^{10.44}~\mathrm{M}_\odot$) as in Section \ref{sec:global_z+m_split} and \citet{jones22}. For both choices of split, we estimate consistent $R_V$ population distributions in the two mass bins. For host galaxies more massive than $10^{10}~\mathrm{M}_\odot$, we estimate an $R_V$ population distribution with mean $\mu_R=2.54\pm0.26$, and standard deviation $\sigma_R=0.81\pm0.32$. For hosts less massive than $10^{10}~\mathrm{M}_\odot$, we estimate $\mu_R=2.79\pm0.58$, and place upper limits on $\sigma_R<1.33~(2.94)$ with 68 (95) per cent posterior probability. For the split at $10^{10.44}~\mathrm{M}_\odot$, we estimate $\mu_R=2.27\pm0.32$ for high mass hosts, and $\mu_R=2.82\pm0.31$ for low mass hosts. In higher mass hosts, we place an upper limit on $\sigma_R<0.78~(1.62)$ at the 68 (95) per cent level. In the lower mass hosts, we estimate $\sigma_R=0.87\pm0.43$, albeit with significant posterior probability density near zero (see Fig.~\ref{fig:raisin_pop_mass_split}).

\begin{figure}
    \centering
    \includegraphics[width=\linewidth]{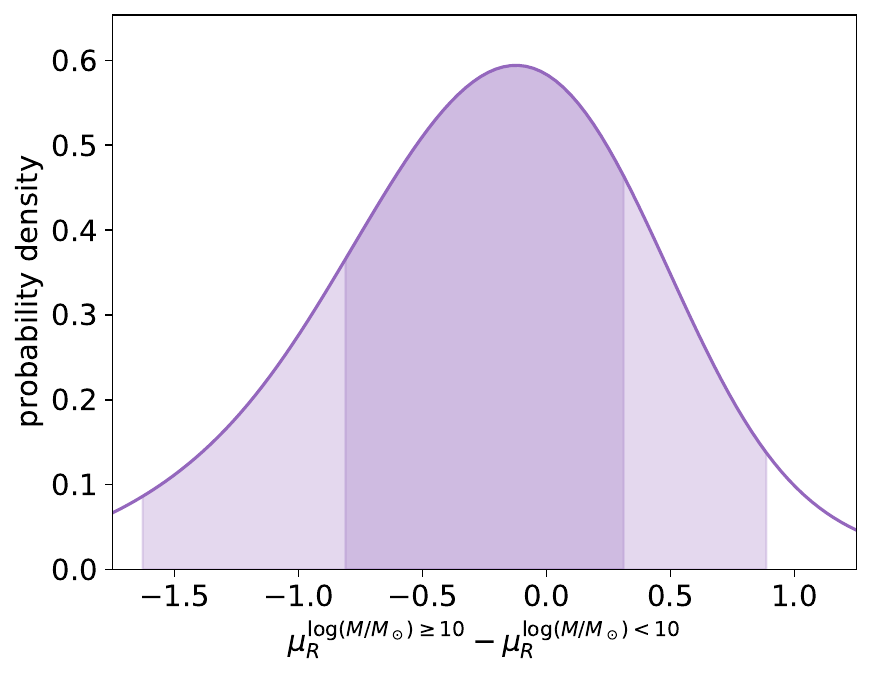}
    \caption{Posterior of low- vs.\ high-mass $\Delta\mu_R^M$, for the CSP+RAISIN combined sample split at $10^{10}~\mathrm{M}_\odot$.}
    \label{fig:delta_muR_mass_split}
\end{figure}

As an alternative to Figure \ref{fig:delta_muR_z_split}, we can use the results from this Section to estimate the posterior distribution of a high- vs.\ low-mass change in $\mu_R$ ($\Delta\mu_R^M=\mu_R^{\text{high-}M}-\mu_R^{\text{low-}M}$). Figure \ref{fig:delta_muR_mass_split} shows this for a mass split at $10^{10}~\mathrm{M}_\odot$. We can see that this shows a slight tendency towards a negative $\Delta\mu_R^M$ (i.e.\ a weak preference of higher $\mu_R$ in lower-mass host galaxies). However, a $\Delta\mu_R^M$ of zero (i.e.\ no difference) is well within the 68 per cent credible interval. With 68 per cent posterior probability, we estimate that $-0.83<\Delta\mu_R^M<0.34$, and with 95 per cent posterior probability, we estimate $-1.66<\Delta\mu_R^M<0.91$. We do not include a Figure for the $10^{10.44}~\mathrm{M}_\odot$ split, but the picture is very similar, albeit with a slightly stronger tendency towards negative $\Delta\mu_R^M$. For this split, we estimate $-0.97<\Delta\mu_R^M<-0.12$ at the 68 per cent level, and $-1.46<\Delta\mu_R^M<0.35$ at the 95 per cent level.

\section{Conclusions}
\label{sec:conclusions}
In this paper, we have used our \textsc{BayeSN} hierarchical model to analyse high-$z$ SN Ia light curves from the RAISIN Survey \citep{jones22}, along with a complementary set of low-$z$ SNe Ia from CSP. We have presented constraints on the dust law $R_V$ (and its population distribution) in SN Ia host galaxies for this redshift range ($0.22\leq z \leq 0.61$). This is the first such analysis using RAISIN data, and our results are the first SN Ia host galaxy $R_V$ constraints at high-$z$ to be based on rest-frame optical and NIR data.

Assuming a single common dust law can describe the RAISIN sample, we estimate $R_V=2.37\pm0.39$. This is fully consistent with the value of $R_V=2.52\pm0.11$ that we estimate for the low-$z$ CSP sample (which, in turn, is fully consistent with our low-$z$ $R_V$ estimates from previous analyses; \citealp{thorp21, thorp22, ward23, grayling24}). We use these results to place probabilistic limits on the shift in $R_V$ between low- and high-redshift ($\Delta R_V^z = R_V^\text{RAISIN} - R_V^\text{CSP}$), estimating that $-0.84<\Delta R_V^z<0.71$, with 95 per cent posterior probability. This constraint does not rely on assuming that $R_V$ is associated with a particular galaxy type/property, and propagating that indirectly to a constraint on $\Delta R_V^z$ based on a model for galaxy evolution across cosmic time. Rather, it a direct constraint of $\Delta R_V^z$, conditional on the data. This has the advantage of depending on fewer astrophysical assumptions, and does not assume that an estimated low-$z$ correlation between $R_V$ and host properties perseveres at high-$z$. However, a direct $\Delta R_V^z$ constraint cannot be easily projected beyond the redshift range of the data that it is obtained from (c.f.\ the progenitor-based mass step model of \citealp{rigault13} or \citealp{childress14}; or the linear $\mu_R$ vs.\ redshift relation of \citealp{grayling24}). Additionally, the utility of a direct constraint requires the high- and low-$z$ samples analysed to be representative (in terms of dust properties, at least) of the broader population of SN Ia hosts in their respective redshift ranges. 

With CSP-I leaning heavily towards the follow-up of SNe from ``targeted'' searches (see \citealp{krisciunas17} \S2, and references therein), the low-$z$ component of our sample will not have a host galaxy distribution that is representative of low-$z$ SN hosts more generally. It is particularly biased in terms of host galaxy mass -- strongly favouring more massive hosts as can be seen from Figure \ref{fig:AV_vs_mass}. This means that the difference in mass distribution between the low- and high-$z$ samples we have analysed is likely more significant than the difference between the underlying SN Ia host mass distribution in the two redshift ranges. \citet{pan14} and \citet{smith20} present detailed discussion of the SN Ia host mass distributions for different surveys and redshift ranges. \citet{pan14} find that the untargetted low-$z$ Palomar Transient Factory SN Ia sample has a less pronounced tendency towards high host masses than other low-$z$ samples (explained by galaxy-targetted surveys favouring bright galaxies), but that the mass distribution still skews higher than for the higher-$z$ SNLS sample. They find that although the galaxy stellar mass function does not strongly evolve with redshift at $z<1$ \citep{drory09}, the convolution of the mass function with a mass--age relation \citep{gallazzi05} and the SN Ia delay time distribution \citep{maoz12} can explain the higher fraction of low-mass hosts at higher-$z$ \citep[see also discussion in][]{childress14}. \citet{smith20} present a comparison of five different surveys; they support the conclusions of \citet{pan14}, but also add that spectroscopic targeting strategy of DES likely favoured faint low-mass hosts (as it is advantageous to get redshifts of these systems whilst the SN is bright; see also \citealp{smith20_overview}). This selection may also effect the RAISIN2 sample we have used here. Nevertheless, our previous work has yielded consistent inferences about $R_V$ when using the CSP data \citep{thorp22}, and when using a more ``untargeted'' low-$z$ sample with a less skewed mass distribution (\citealp{thorp21}; using data from the Foundation Supernova Survey; \citealp{foley18, jones19}), suggesting that the low-$z$ selection effects may not be a major cause for concern for $R_V$ inference. 

Due to the caveats associated with our estimated limits on $\Delta R_V^z$, we will forgo a detailed propagation of these limits into an estimate of potential cosmological bias. Such an effort will be better deferred to a future study incorporating a treatment of selection effects. This could be achieved via a survey-simulation-based approach using SNANA \citep{kessler09_snana} as in the RAISIN cosmology analysis \citep{jones22}, by integrating a Bayesian selection effect treatment (\`a la UNITY; \citealp{rubin15, rubin23}; or BAHAMAS; \citealp{shariff16, rahman22}) into our own hierarchical model, or via simulation based inference \citep[e.g.][]{karchev23, karchev24}. Independent of any implications for cosmology, however, constraints on dust from SNe Ia can still provide a unique insight into dust in galaxies beyond the Milky Way, complementary to the information provided by different probes. As pointed out by \citet{keel23}, SNe Ia (being distributed fairly uniformly throughout their hosts; e.g. \citealp{galbany14}) can give valuable information about the dust further from bright or star forming regions, and they provide information about the dust extinction along specific sightlines (rather than the integrated effect of dust over a large region or the whole galaxy).

As well as estimating the single best-fitting $R_V$ values in the low- and high-$z$ samples, we also perform an analysis where the assumption of zero variance is relaxed, and the parameters (mean, $\mu_R$, and standard deviation, $\sigma_R$) of a Gaussian $R_V$ population distribution are fitted. For the low-$z$ CSP sample we estimate $\mu_R=2.62\pm0.26$ and $\sigma_R=0.82\pm0.37$, whilst for the high-$z$ RAISIN sample we estimate $\mu_R=2.58\pm0.57$ and place an upper limit on $\sigma_R<0.90~(2.42)$ with 68 (95) per cent posterior probability. Our population distribution estimate at low-$z$ is consistent with our previous analyses \citep{thorp21, thorp22}, and the estimates at high- and low-$z$ are consistent to within their uncertainties. As with the single-$R_V$ analysis, we are able to place limits on the potential redshift drift of the \textit{mean} $R_V$ ($\Delta \mu_R^z = \mu_R^\text{RAISIN} - \mu_R^\text{CSP}$), finding that $-1.16<\Delta\mu_R^z<1.38$ with 95 per cent posterior probability. The credible interval here is slightly wider than the equivalent $\Delta R_V^z$ posterior, and permits a wide range of possibilities. Analysis of a large optical sample \citep{grayling24} has yielded similar constraints. Nevertheless, it allows us to confidently rule out very extreme drifts in mean $R_V$ of $\gtrsim1.5$, and is a promising preliminary result that should be built on in future analyses.

Finally, given the consistency of our $R_V$ population distribution inferences in the CSP and RAISIN samples, we use the combined CSP+RAISIN data to investigate the mass-dependence of the host galaxy $R_V$ distribution. In combination, these data have a fairly even split between host masses above and below $10^{10}~\mathrm{M}_\odot$ ($46/79$ above, $33/79$ below), albeit with the quirk that the high-mass subset is dominated by CSP ($35/46$), whilst the low-mass subset is dominated by RAISIN ($26/33$). For host galaxies more massive than $10^{10}~\mathrm{M}_\odot$, we estimate an $R_V$ population distribution with mean $\mu_R=2.54\pm0.26$, and standard deviation $\sigma_R=0.81\pm0.32$. For hosts less massive than $10^{10}~\mathrm{M}_\odot$, we estimate $\mu_R=2.79\pm0.58$, and $\sigma_R<1.33~(2.94)$ at the 68 (95) per cent level. Our results can be propagated to a constraint on the high- minus low-mass difference in $\mu_R$ ($\Delta \mu_R^M = \mu_R^{\text{high-}M} - \mu_R^{\text{low-}M}$), leading to a 95 per cent credible interval encompassing $-1.66<\Delta\mu_R^M<0.91$. For an alternative sample split at $10^{10.44}~\mathrm{M}_\odot$, the results are similar -- i.e.\ no strong exclusion of $\Delta\mu_R^M=0$.

In summary, we have utilised the unique high-$z$ rest-frame NIR data provided by the RAISIN Survey to investigate the mass- and redshift-dependence of the dust law $R_V$ in SN Ia host galaxies. Regarding the potential redshift-evolution of $R_V$ (independent of other host galaxy properties), we have placed direct constraints on the difference in (mean) $R_V$ between SN Ia hosts at $z<0.1$ and $0.2\lesssim z\lesssim 0.6$. These are the first constraints of this nature to make use of rest-frame NIR data across the full redshift range studied. We do not find statistically significant evidence for a non-zero shift in mean $R_V$ between low- and high-$z$, and are able to rule out  an absolute shift in mean $R_V$ of $\gtrsim1.5$ with high confidence. These results complement the optical analysis of \citet{grayling24}, who estimate a $\mu_R$ vs.\ $z$ correlation coefficient of $\eta_R=-0.38\pm0.70$. Given that a shift in $R_V$ or the colour--luminosity coefficient $\beta$ by $\approx1.4$ between low- and high-$z$ could easily propagate to a significant bias in $w$, future work is needed to control this potential systematic. One interesting feature of our current results, and those of \citet{grayling24} is the weak preference for a decrease in $R_V$ with redshift. Many recent analyses have suggested a similarly weak preference for lower mean $R_V$ in more massive galaxies \citep[e.g.][]{brout21, thorp21, johansson21, thorp22, popovic23, grayling24}. However, lower-redshift samples of supernovae tend to be weighted towards more massive host galaxies \citep[see e.g.][]{pan14, smith20} -- something that is also true for the RAISIN sample in comparison to CSP (Fig.\ \ref{fig:AV_vs_mass}). The preference seen for higher $R_V$ at lower redshifts is thus slightly surprising.

The analysis we have presented here lays the groundwork for future NIR investigations, and can be developed further via the incorporation of a robust Bayesian treatment of selection effects \citep[see e.g.][]{rubin15, shariff16, rahman22, rubin23, karchev23, karchev24}. Investigation of intrinsic colour evolution with redshift would also be a worthwhile avenue to pursue. The promising results we have obtained using a sample of 37 high-$z$ SNe Ia illustrate the value of having rest-frame NIR data at higher redshifts, adding to the case already made by the RAISIN cosmology results \citep{jones22}. Such data could potentially be obtained by the \textit{Roman Space Telescope} HLTDS \citep{hounsell18, rose21}, although the current reference survey design prioritises the rest-frame optical \citep{rose21}. A design more focused on the rest frame NIR could be hugely valuable, and would be a great complement to the rest-frame optical data obtained by LSST (for further discussion of \textit{Roman}--LSST synergies, see \citealp{foley18_roman, rhodes19, rose21_synergy, bianco24}). The discovery and analysis of SN Ia siblings (multiple SNe Ia in the same host; e.g.\ \citealp{elias81, hamuy91, stritzinger10, gall18, burns20, scolnic20, graham22, kelsey23, dwomoh23}) at high-$z$ will also prove complimentary, with these systems offering greater leverage for constraining dust properties and colour--luminosity correlations than galaxies that host a lone optically observed supernova \citep{biswas22, ward22}. In fact, \citet{ward22} liken the leverage gained from the presence of a sibling SN Ia to the benefit gained from having NIR data. Understanding any redshift-dependent systematics/biases associated with host galaxy dust will be of critical importance to the LSST and \textit{Roman} dark energy analyses. And the impact of these effects is already being felt in current analyses \citep{vincenzi24, des24}. The continued work by the community to observe \citep[e.g.][]{woodvasey08, friedman15, krisciunas17, phillips19, johansson21, konchady22, mullerbravo22, jones22, peterson23} and model \citep[e.g.][]{mandel09, mandel11, burns11, burns14, pierel18, avelino19, mandel22, pierel22, grayling24} SNe Ia in the NIR will be crucial to furthering our understanding and controlling dust-related systematics.

%%%%%%%%%%%%%%%%%%%%%%%%%%%%%%%%%%%%%%%%%%%%%%%%%%
\section*{Acknowledgements}
ST thanks Matt Auger--Williams and Roberto Trotta for useful feedback and discussions of this work during his PhD defense. We thank Joel Johansson and the anonymous referee for useful comments.

ST was supported by the Cambridge Centre for Doctoral Training in Data-Intensive Science funded by the UK Science and Technology Facilities Council (STFC), and in part by the European Research Council (ERC) under the European Union’s Horizon 2020 research and innovation programme (grant agreement no.\ 101018897 CosmicExplorer). KSM acknowledges funding from the European Research Council under the European Union’s Horizon 2020 research and innovation programme (Grant Agreement No. 101002652). This project has been made possible through the ASTROSTAT-II collaboration, enabled by the Horizon 2020, EU Grant Agreement No. 873089.  

This work made use of the Illinois Campus Cluster, a computing resource that is operated by the Illinois Campus Cluster Program (ICCP) in conjunction with the National Center for Supercomputing Applications (NCSA) and which is supported by funds from the University of Illinois at Urbana-Champaign.

\section*{Data Availability}
%The inclusion of a Data Availability Statement is a requirement for articles published in MNRAS. Data Availability Statements provide a standardised format for readers to understand the availability of data underlying the research results described in the article. The statement may refer to original data generated in the course of the study or to third-party data analysed in the article. The statement should describe and provide means of access, where possible, by linking to the data or providing the required accession numbers for the relevant databases or DOIs.
All data used are publicly available. The Carnegie Supernova Project data are presented in \citet{krisciunas17}. The data from the RAISIN Survey are presented in \citet{jones22}. The RAISIN data release containing all relevant light curves is available publicly on GitHub\footnote{\url{https://github.com/djones1040/RAISIN_DataRelease}}. A new \textsc{BayeSN} code \citep{grayling24} is also available on GitHub\footnote{\url{https://github.com/bayesn/bayesn}} and will be integrated into SNANA\footnote{\url{https://github.com/RickKessler/SNANA}} \citep{kessler09_snana}. 

%%%%%%%%%%%%%%%%%%%% REFERENCES %%%%%%%%%%%%%%%%%%

% The best way to enter references is to use BibTeX:

\bibliographystyle{mnras}
\bibliography{main} % if your bibtex file is called example.bib

%%%%%%%%%%%%%%%%%%%%%%%%%%%%%%%%%%%%%%%%%%%%%%%%%%

%%%%%%%%%%%%%%%%% APPENDICES %%%%%%%%%%%%%%%%%%%%%

%Appendices go here

\appendix

\section{SN Ia Dust Laws Since 2020}
\label{app:rv}
Over the past three years, there has been a proliferation of analyses that have said something about the $R_V$ distribution in SN Ia host galaxies. In Table \ref{tab:literature}, we attempt to summarise as many key results as possible from this time period. We also include several key earlier results from 2010--2020 that are frequently used as a reference for more recent measurements (but we omit earlier landmark results in the history of SN Ia $R_V$ estimation; e.g.\ \citealp{joeveer82, riess96, krisciunas00, krisciunas07}). We confine this comparison primarily to studies of samples (as opposed to single SNe), and do not quote every result from each paper in cases where there are many analysis variants. For each study, we state (where applicable) reported values for the following quantities: $R_V$ or $\mu(R_V)$ for the full sample; $\sigma(R_V)$ for the full sample; $\mu(R_V)$ and $\sigma(R_V)$ either side of the ``classic'' $10^{10}~\mathrm{M}_\odot$ mass step. We note that many of the listed papers presented more analysis variants than we can easily summarise here.

% Don't change these lines
\bsp	% typesetting comment
\label{lastpage}

\begin{landscape}
\begin{table}
    \centering
    \begin{threeparttable}
        \caption{Literature $R_V$ estimates for populations of SN Ia hosts since 2010.}
        \label{tab:literature}
        \begin{tabular}{l c c c c c c c c c c}\toprule
             Paper & Model\tnote{a} & Data Type\tnote{b} & Sample & $R_V$\tnote{c} & $\mu(R_V)$\tnote{d} & $\sigma(R_V)$\tnote{e} & $\mu(R_V)|$HM\tnote{f} & $\sigma(R_V)|$HM\tnote{f} & $\mu(R_V)|$LM\tnote{g} & $\sigma(R_V)|$LM\tnote{g}\\\midrule
             \citet{folatelli10}\tnote{\dh} & \textsc{SNooPy}\tnote{\S} & Phot.\ & CSP\tnote{3} & $3.20\pm0.40$ & - & - & - & - & - & - \\
             \citet{chotard11}\tnote{\dh} & Other\tnote{$\nabla$} & Spec.\ & SNfactory\tnote{1} & $2.78\pm0.34$ & - & - & - & - & - & - \\
             \citet{mandel11}\tnote{\dh} & \textsc{BayeSN} & Phot.\ & CfA\tnote{2}, CSP\tnote{3,10} & $1.60\pm0.10$ & $2.30\pm0.30$ & - & - & - & - & -\\
             \citet{burns14}\tnote{\"{o},\dh} & \textsc{SNooPy} & Phot.\ & CSP\tnote{3} & $1.76\pm0.18$ & $2.50\pm0.30$ & $0.88\pm0.05$ & - & - & - & -\\
             \citet{burns14}\tnote{\"{o},\th} & \textsc{SNooPy} & Phot.\ & CSP\tnote{3} & $2.15\pm0.16$ & $2.30\pm0.30$ & $0.84\pm0.05$ & - & - & - & - \\
             \citet{mandel17} & SALT2 & Phot.\ & SuperCal\tnote{19} & $2.76\pm0.34$ & - & - & - & - & - & -\\
             \citet{leget20}\tnote{\dh} & SUGAR & Spec. & SNfactory\tnote{1} & 2.60 & - & - & - & - & - & -\\
             \citet{brout21}\tnote{\aa} & SALT2\tnote{$\dag$} & Phot. & Various\tnote{2--8} & - & $2.00\pm0.20$ & $1.40\pm0.20$ & $1.50\pm0.25$ & $1.30\pm0.20$ & $2.75\pm0.35$ & $1.30\pm0.20$\\
             \citet{thorp21}\tnote{\th} & \textsc{BayeSN}\tnote{} & Phot. & Foundation\tnote{4} & $2.61\pm0.21$ & $2.70\pm0.25$ & 0.37 (0.61) & $2.72\pm0.23$ & 0.32 (0.62) & $3.49\pm0.61$ & 0.89 (1.87)\\
             \citet{johansson21}\tnote{\dh} & \textsc{SNooPy} & Phot. & Various NIR\textsuperscript{2,3,9--12} & - & 1.90 & $0.90$ & 1.70 & 0.80 & 2.20 & 0.90\\
             \citet{arima21} & Multiband Stretch\tnote{*} & Phot. & Low-$z$\tnote{13} & $3.00^{+0.40}_{-0.30}$ & - & - & - & - & - & -\\
             \citet{mandel22}\tnote{\th} & \textsc{BayeSN} & Phot. & Low-$z$ NIR\tnote{14} & $2.89\pm0.20$ & - & - & - & - & - & -\\
             \citet{thorp22}\tnote{\th} & \textsc{BayeSN} & Phot. & CSP\tnote{3} & - & $2.59\pm0.14$ & $0.62\pm0.16$ & $2.54\pm0.15$ & $0.56\pm0.18$ & $2.78\pm0.70$ & 1.62 (3.01)\\
             \citet{wiseman22}\tnote{\aa} & SALT2 & Phot. & DES5YR\tnote{15} & - & - & - & 1.75 & 1.00 & 3.00 & 1.00\\
             \citet{popovic23}\tnote{\"{a}} & SALT2 & Phot. & Pantheon+\tnote{16} & - & - & - & $2.14\pm0.25$ & $1.06\pm0.43$ & $3.03\pm0.38$ & $1.48\pm0.42$\\
             \citet{ward23}\tnote{\th} & GP\tnote{$\infty$} & Phot. & Low-$z$ NIR\textsuperscript{2,3,9,10} & - & $2.61^{+0.38}_{-0.35}$ & 0.92 (1.96) & - & - & - & -\\
             \citet{meldorf22}\tnote{\"{o},\ss} & \textsc{Bagpipes}\tnote{||} & Gal.\ Phot. & DES5YR\tnote{15,17} & - & - & - & $2.61\pm0.07$ & $0.86\pm0.06$ & $3.02\pm0.05$ & $0.82\pm0.04$\\
             \citet{duarte22}\tnote{\"u} & \textsc{Prospector}\tnote{\P} & Gal.\ Phot. & DES3YR\tnote{5,18} & - & - & - & $2.67\pm0.11$ & 0.97 & $3.09\pm0.13$ & 1.16\\
             \citet{smadja23}\tnote{\th} & Other\tnote{$\nabla$} & Spec., Phot. & SNfactory\tnote{1} & $2.22\pm0.17$ & - & - & - & - & - & -\\
             \citet{wojtak23} & SALT2 & Phot. & SuperCal\tnote{19} & - & $3.13^{+0.65}_{-0.59}$ & $0.95^{+0.31}_{-0.28}$ & - & - & - & -\\
             \citet{karchev23_sim, karchev24}\tnote{\ae,\th} & \textsc{BayeSN} & Phot. & CSP\tnote{3} & - & - & - & - & - & - & -\\
             \citet{vincenzi24}\tnote{\"{a}} & SALT3\tnote{$\ddag$} & Phot. & DES5YR\tnote{15} & - & - & - & 1.66 & 0.95 & 3.25 & 0.93\\
             \citet{grayling24}\tnote{\th} & \textsc{BayeSN} & Phot. & Various\tnote{4--6} & - & $2.58\pm0.14$ & $0.59\pm0.20$ & $2.51\pm0.16$ & $0.47\pm0.21$ & $2.74\pm0.35$ & $0.93\pm0.32$\\
             \citet{grayling24}\tnote{\o, \th} & \textsc{BayeSN} & Phot. & Various\tnote{4--6} & - & - & - & $2.26\pm0.14$ & 0.27 (0.49) & $3.36\pm0.51$ & 1.07 (1.81)\\
             \midrule
             Thorp et al.\ (this work)\tnote{\th} & \textsc{BayeSN} & Phot. & CSP\tnote{3} & $2.52\pm0.11$ & $2.62\pm0.26$ & $0.82\pm0.37$ & - & - & - & -\\
             &&& RAISIN\tnote{20} & $2.37\pm0.39$ & $2.58\pm0.57$ & 0.90 (2.42) & - & - & - & -\\
             &&& CSP\tnote{3},\, RAISIN\tnote{20} & - & - & - & $2.54\pm0.26$ & 0.81 (0.32) & $2.79\pm0.58$  & 1.33 (2.94)\\
             \bottomrule
        \end{tabular}
        \begin{tablenotes}
            \setlength{\multicolsep}{0cm}
            \begin{multicols}{3}
            %\begin{minipage}[t]{0.4\linewidth}
            \item [a] Primary light curve model (or galaxy SED model, when applicable) used for the data
            \item [b] Primary data analysed: Spec.\ = SN spectroscopy; Phot.\ = SN photometry; Gal.\ Phot.\ = Galaxy photometry
            \item [c] Quoted estimate of a single sample-level $R_V$
            \item [d] Quoted estimate of a population mean $R_V$
            \item [e] Quoted estimate for $R_V$ population dispersion; values formatted $X$ ($Y$) are 68\% (95\%) upper bounds
            \item [f] Estimate for $R_V$ population mean and std.\ dev.\ for $\geq10^{10}~\mathrm{M}_\odot$
            \item [g] Estimate for $R_V$ population mean and std.\ dev.\ for $<10^{10}~\mathrm{M}_\odot$

            \item[\dh] These analyses uses the \citet{cardelli89} or \citet{odonnell94} dust law
            \item[\th] These analyses uses the \citet{fitzpatrick99} dust law
            \item[\ss] This analysis uses the \citet{salim18} modification of the \citet{calzetti94} attenuation law
            \item[\"u] This analysis uses the \citet{calzetti00} attenuation law (modified by \citealp{noll09, kriek13})

            \item [\aa] These analyses searched over a finite grid of models
            \item [\"{a}] These analyses analysed their SALT fits using \textsc{Dust2Dust} \citep{popovic23}
            \item [\"{o}] These analyses fit a two component Gaussian mixture; we quote the mean for the dominant component containing $\gtrsim90\%$ of SNe
            \item [\ae] \citeauthor{karchev23_sim} present estimates, but not numerical summaries
            \item [\o] For this analysis, \citeauthor{grayling24} allow  a mass dependent difference in the intrinsic SN Ia SED (i.e.\ a ``non-gray'' mass step)

            \item [$\infty$] \citeauthor{ward23} use a Gaussian process to estimate peak colours
            \item [$\nabla$] These analyses use bespoke models based on equivalent widths
            \item [$\dag$] \citet{guy07}
            \item [\S] \citet{burns11}
            \item [*] \citet{takanashi08, takanashi17}
            \item [||] \citet{carnall18}
            \item [\P] \citet{leja17}
            \item [$\ddag$] \citet{kenworthy21, taylor23}
            %\end{minipage}%
            %\begin{minipage}[t]{0.25\linewidth}
            \item [1] \citet{aldering20}
            \item [2] \citet{hicken09, hicken12}
            \item [3] \citet{krisciunas17}
            \item [4] \citet{foley18, jones19}
            \item [5] \citet{brout19_des}
            \item [6] \citet{rest14, scolnic18}
            \item [7] \citet{betoule14}
            \item [8] \citet{sako18}
            \item [9] \citet{johansson21}
            \item [10] \citet{woodvasey08, friedman15}
            \item [11] \citet{baronenugent12, stanishev18}
            \item [12] \citet{amanullah15}
            \item [13] \citet{takanashi08}
            \item [14] \citet{avelino19}
            \item [15] \citet{vincenzi24}
            \item [16] \citet{scolnic22}
            \item [17] \citet{hartley22}
            \item [18] \citet{kelsey21}
            \item [19] \citet{scolnic15}
            \item [20] \citet{jones22}
            %\end{minipage}
            \end{multicols}
        \end{tablenotes}
    \end{threeparttable}
\end{table}
\end{landscape}

%%%%%%%%%%%%%%%%%%%%%%%%%%%%%%%%%%%%%%%%%%%%%%%%%%

\end{document}